\documentclass[twocolumn]{IEEEtran}

\usepackage{amsmath}
\usepackage{amsfonts}
\usepackage{subfigure}
\usepackage{mathrsfs}
\usepackage{pifont}
\usepackage{amssymb}
\usepackage{verbatim}
\usepackage{upgreek}
\usepackage{color}
\usepackage[final]{graphicx}
\usepackage{epsfig}
\usepackage{booktabs}
\usepackage{bm}
\usepackage{setspace}
\usepackage{url}
\usepackage[latin1]{inputenc}

\newcommand{\bzero}{\mbox{\boldmath{$0$}}}
\newcommand{\bA}{\mbox{\boldmath{$A$}}}
\newcommand{\ba}{\mbox{\boldmath{$a$}}}

\newcommand{\bI}{\mbox{\boldmath{$I$}}}
\newcommand{\bM}{\mbox{\boldmath{$M$}}}
\newcommand{\bn}{\mbox{\boldmath{$n$}}}
\newcommand{\bR}{\mbox{\boldmath{$R$}}}
\newcommand{\br}{\mbox{\boldmath{$r$}}}
\newcommand{\bS}{\mbox{\boldmath{$S$}}}
\newcommand{\bs}{\mbox{\boldmath{$s$}}}
\newcommand{\bU}{\mbox{\boldmath{$U$}}}

\newcommand{\bv}{\mbox{\boldmath{$v$}}}

\newcommand{\bw}{\mbox{\boldmath{$w$}}}
\newcommand{\bX}{\mbox{\boldmath{$X$}}}
\newcommand{\bx}{\mbox{\boldmath{$x$}}}
\newcommand{\by}{\mbox{\boldmath{$y$}}}

\newcommand{\bSigma}{\mbox{\boldmath{$\Sigma$}}}
\newcommand{\blambda}{\mbox{\boldmath{$\lambda$}}}
\newcommand{\bLambda}{\mbox{\boldmath{$\Lambda$}}}
\newcommand{\tr}{\mbox{\rm tr}\, }
\newcommand{\diag}{\mbox{\boldmath\bf diag}\, }
\def\E{{\mathbb E}}
\newtheorem{theorem}{Theorem}[section]

\newtheorem{lemma}[theorem]{Lemma}

\title{A Geometric Approach to Covariance Matrix Estimation and its Applications to Radar Problems}

\author{A. Aubry, \emph{Senior Member, IEEE}, A. De Maio  \emph{Fellow, IEEE}, and L. Pallotta, \emph{Member, IEEE}\thanks{A. Aubry and A. De Maio (corresponding author) are with Universit\`a degli Studi di Napoli ``Federico II'', Dipartimento di Ingegneria Elettrica e delle Tecnologie dell'Informazione, Via Claudio 21, I-80125 Napoli, Italy. E-mail: augusto.aubry@unina.it, ademaio@unina.it.}
\thanks{L. Pallotta is with CNIT, viale G.P. Usberti, n. 181/A - 43124 Parma, c/o udr Universit\`a ``Federico II'', via Claudio 21, I-80125 Napoli, Italy. E-mail: luca.pallotta@unina.it.}}

\begin{document}

\maketitle
\vspace{-0.9cm}
\begin{abstract}
A new class of disturbance covariance matrix estimators for radar signal processing applications is introduced following a geometric paradigm. Each estimator is associated with a given unitary invariant norm and performs the  sample covariance matrix {\em projection}  into a specific set of structured covariance matrices. Regardless of the considered norm, an efficient solution technique to handle the resulting constrained optimization problem is developed. Specifically, it is shown that the new family of distribution-free estimators shares a shrinkage-type form;  
besides, the eigenvalues estimate just requires the solution of a one-dimensional convex problem whose objective function depends on the considered unitary norm. For the two most common norm instances, i.e., Frobenius and spectral, very efficient algorithms are developed to solve the aforementioned one-dimensional optimization leading to almost closed form covariance estimates. At the analysis stage, the performance of the new  estimators is assessed in
terms of achievable Signal to Interference plus Noise Ratio (SINR) both for a spatial and a Doppler processing assuming different data statistical characterizations. The results show that interesting SINR improvements with respect to some counterparts available in the open literature can be achieved especially in training starved regimes.
\end{abstract}


\section{Introduction}
Interference covariance matrix estimation is a longstanding and basic problem in adaptive radar signal processing and naturally arises in several areas such as target detection, direction of arrival estimation, sidelobe cancelling, and secondary data selection \cite{Reed,robey1992cfar,chen1999screening,kelly1986adaptive} (just to list a few). Conventional adaptive architectures (such as Sample Matrix Inversion (SMI) Doppler filter \cite{Reed}, Kelly's receiver \cite{kelly1986adaptive}, and spatial beamformers \cite{Farina_book}) resort to the sample covariance matrix of a secondary data set collected from range gates spatially close to the one under test to estimate the interference covariance. These algorithms are often very prohibitive because they lean on the assumption that the environment remains stationary and homogeneous during the adaptation process. Precisely, they provide satisfactory performance when the secondary vectors share the same spectral properties of the interference in the test cell, are statistical independent, and their number is higher than twice the useful signal dimension \cite{Reed}. These requisites however may represent important limitations since in real environments the number of data where the disturbance is homogeneous (often referred to as sample support) is very limited. Besides, poor training data selection, in such adaptive algorithms, can result in severe radar performance degradation \cite{Armstrong} and \cite{Himed}.\\
A viable means to thwart the lack of a sufficient number of homogeneous secondary data is to capitalize some a-priori information about the radar surrounding environment, namely to realize a knowledge-based/cognitive processing \cite{gini2008knowledge} 
so as to restrict the uncertainty region of the unknown parameters.  According to this processing paradigm, several approaches have been pursued in the open literature assuming different structural models as well as statistical distributions of the data. In particular, both homogeneous and heterogeneous interference environments are dealt with. As to the former scenario, the training data are modeled as independent and identically distributed (i.i.d.), 
zero-mean, circularly symmetric Gaussian vectors. Whereas, the latter context mainly considers clutter power variations within the sample support. Thus, assuming homogeneity, in \cite{Gerlach} the Maximum Likelihood (ML) covariance matrix estimator is derived modeling the disturbance as the sum of a coloured interference plus white disturbance; in \cite{K_m_only}, the ML estimation of an unstructured covariance matrix with a condition number upper bound requirement is considered; in \cite{TSP}, using the same covariance structure as in \cite{Gerlach}, the ML estimator is derived when a constraint on the condition number is imposed too; in \cite{kang2014rank}, a rank-constrained ML estimator is developed; furthermore, relying on a Mean Square Error (MSE) design criterion, in \cite{Ledoit} and \cite{Chen}, some shrinkage estimators are proposed. With reference to heterogeneous scenarios, compound Gaussian statistical models (such as K-distributed or Gamma amplitudes) are usually exploited to account for clutter returns spikiness. In \cite{6035802}, assuming a clutter dominated environment, a unified framework to regularize the ML estimate in scaled Gaussian models (e.g., elliptical distributions, compound-Gaussian processes and spherically invariant random vectors) is developed in order to enhance Tyler's estimator in the presence of a small sample support exploiting a-priori information on the covariance structure. In \cite{7289442}, Tyler's robust covariance M-estimator under group symmetry constraints, such as circulant, persymmetric, and proper quaternion matrices, is considered. Precisely, it is provided an iterative fixed point algorithm to compute the constrained estimate. Finally, in \cite{SunPalomar}, an iterative algorithm to estimate, according to the ML approach, both the clutter subspace and the covariance is proposed, assuming the disturbance composed of a low rank compound Gaussian clutter plus a white Gaussian noise contribution. Further technically sound and effective covariance estimators can be found in \cite{new_3,new_2,new_1,new_9,new_4,de2003maximum}.\\
These mentioned algorithms lean on specific assumptions about the data statistical characterization
and may suffer performance loss in the presence of model mismatches due to for instance quantization effects, constant modulus jamming signals, phase noise \cite{Stimson,phase_noise_1,phase_noise_2}. To overcome this shortcoming and endow robustness
to the estimation process, some strategies derived from geometric considerations on the metric space of the covariance
matrices have been also developed. 
In this respect, in \cite{barbaresco2013information} and \cite{balaji2014information} techniques based on Riemannian $p$-mean (e.g., Fr\'{e}chet median and Karcher barycenter) evaluation are developed to estimate suitably structured interference covariance matrices showing that substantial ameliorations over classical algorithms can be attained. In \cite{barbaresco2013information} and \cite{Barb_2} also an extension of the conventional Ordered Statistic (OS) framework \cite{Rich} is proposed relying on the Riemannian $p$-mean computation of Toepliz or Toeplitz-Block-Toeplitz space-time covariance matrices. Besides, in \cite{aubry2013covariance} and \cite{aubry2014median}, covariance estimates defined through geometric barycenters/medians (associated with specific distances in the space of Hermitian matrices) of structured covariance estimates are exploited both for training data selection and adaptive radar detection highlighting significant gains with respect to the classic sample covariance matrix. Finally, in \cite{new_5,new_10,new_6,new_7,new_8,dryden2009non} other interesting geometric-inspired procedures are devised.\\
In this paper, leveraging on a geometric criterion, a novel class of covariance  estimators that do not consider any assumption on the statistical characterization of the secondary data is proposed and analyzed. Each estimator is associated with a given unitary invariant norm and performs the sample covariance matrix projection into a specific set of structured covariance matrices. Precisely, this set encompasses the matrices modeled as the sum of an unknown positive semi-definite matrix (describing coloured interference and clutter) plus a term proportional to the identity matrix (related to white disturbance). Besides, a constraint on the condition number is accounted for so as to control the numerical stability of the resulting adaptive algorithms \cite{Farina_book}. Regardless of the considered norm, an efficient solution technique to handle the formulated constrained optimization problem is developed. Precisely, each estimator exhibits a shrinkage-type form and its evaluation requires the sample covariance matrix spectral decomposition as well as the solution of a one-dimensional convex problem whose objective function depends on the considered unitary norm. \\
At the analysis stage, the performance of the new class of distribution-free estimators is evaluated in terms of achievable Signal-to-Interference-plus-Noise Ratio (SINR) for
different sample support sizes, assuming both a spatial and a Doppler processing scenario. The results show that interesting SINR improvements can be achieved with respect to some counterparts available in the open literature also with a reduced computational complexity.\\
The remainder of this paper is organized as follows. Section \ref{formulazione_problema} is devoted to the description of the system model as well as the formulation of the covariance matrix estimation problem. In Section \ref{stimatori_matrice}, an efficient procedure to solve the resulting constrained optimization is developed. In Section \ref{risultati}, the performance of the proposed distribution-free estimators is assessed. Finally, Section \ref{conclusioni} concludes the paper and provides some possible future research tracks.

\section*{Notation}

We adopt the notation of using boldface for vectors $\ba$ (lower case), and matrices $\bA$ (upper case). The transpose and the conjugate transpose operators are denoted by the symbols $(\cdot)^T$ and $(\cdot)^\dag$ respectively. $\tr\{\cdot\}$ is the trace of the square matrix argument. $\bI$ and ${\bf 0}$ denote respectively the identity matrix and the matrix with zero entries (their size is determined from the context). $\diag(\ba)$ indicates the diagonal matrix whose $i$-th diagonal element is the $i$-th entry of $\ba$.  ${\mathbb{R}}^N$, ${\mathbb{C}}^N$, ${\mathbb{C}}^{N,K}$, and ${\mathbb{H}}^N$ are respectively the sets of $N$-dimensional vectors of real numbers, of $N$-dimensional vectors of complex numbers, of $N\times K$ matrices of complex numbers, and of $N\times N$ Hermitian matrices. 
The curled inequality symbol $\succeq$ (and its strict form $\succ$) is used to denote generalized matrix inequality: for any $\bA\in{\mathbb{H}}^N$, $\bA\succeq\bzero$ means that $\bA$ is a positive semi-definite matrix ($\bA\succ\bzero$ for positive definiteness). $\|\cdot\|$ denotes an arbitrary unitary invariant matrix norm operator, while the specific spectral and Frobenius instances are indicated by $\|\cdot\|_2$ and $\|\cdot\|_F$, respectively. The letter $j$ represents the imaginary unit (i.e. $j=\sqrt{-1}$). For any complex number $x$, $|x|$ represents the modulus of $x$. Finally, $\E\left[\cdot\right]$ denotes statistical expectation and for any optimization problem ${\cal{P}}$ $v({\cal{P}})$ represents its optimal value. 

\section{Problem Formulation}\label{formulazione_problema}
In this section, the problem of estimating the covariance matrix $\bM \in \mathbb{H}^N$ of $K$ secondary data $\br_1, \ldots, \br_K$  modeled as $N$-dimensional, circularly symmetric, zero-mean random vectors, is addressed. It is assumed that these vectors  share the same second order statistical characterization, i.e.,
\begin{equation}
\E[\br_i \br_i^\dag]= \bM, \quad \text{for } i=1,\ldots, K,
\end{equation}
but are drawn from an arbitrary and unknown joint probability distribution. 

\begin{figure}[h!]
\begin{center}
\includegraphics[width=0.4\textwidth]{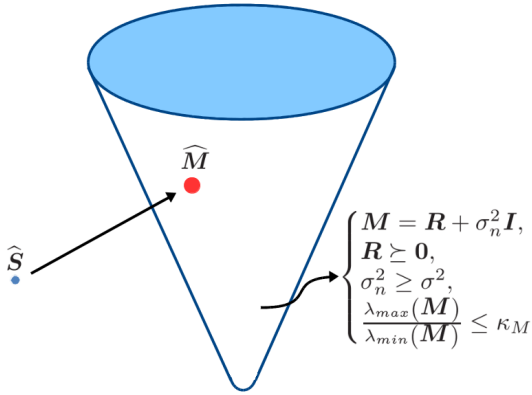}
\end{center}
\caption{Projection of the sample covariance matrix into a specific set through a unitary invariant matrix norm.\label{fig_projection}}
\end{figure}

According to the previous assumptions, classic estimation approaches  such as the ML or the minimum MSE strategies can be no longer pursued since they need the data statistical distribution  knowledge. Hence, in this work a new family of covariance estimators based on geometric considerations 
is introduced. Specifically, the idea is to estimate the data covariance matrix performing the projection\footnote{Notice that, the projection operator is usually defined assuming the reference set being convex and closed as well as the norm being induced by an inner product. Nevertheless, with a slight abuse of notation, in this paper we continue to define as projection the point $\bx_p$ minimizing the  distance (based on a specific metric) between a given point $\bx$ and the reference set as long as the point $\bx_p$ can be uniquely identified.}, as induced by a specific metric, of the sample covariance matrix 

\begin{equation}
\widehat{\bS}=\frac{1}{K}\sum_{i=1}^K \br_i \br_i^\dag,
\end{equation}
into the set 
\begin{equation}\label{vincoli}
\left\{
\begin{array}[c]{ll}
&\bM = \bR + \sigma_n^2\bI, \\
&\bR\succeq {\bf 0},\\
&\sigma_n^2\geq \sigma^2,\\
&\frac{\lambda_{max}\left(\bM\right)}{\lambda_{min}\left(\bM\right)}\leq \kappa_M,
\end{array}
\right.
\end{equation}
where $\sigma^2>0$ is a lower bound to the white interference power, $\bR$ accounts for the colored interfering contribution, and $\kappa_M\geq 1$ is an upper bound to the covariance condition number. This uncertainty set accounts for an interference covariance  structure that is commonly met in adaptive radar signal processing applications and ensures a well conditioned estimate necessary to compute the adaptive radar weight vector. The effectiveness of this model has already been proved in \cite{TSP}, where the training data are assumed i.i.d., 
zero-mean, circularly symmetric Gaussian vectors and the ML estimate is derived. Hence,
the main goal of this work is the development of a distribution-free approach which provides a robust alternative to \cite{TSP} when inference on the interference statistics is not possible. 
A pictorial representation of the geometric-based estimation process is reported in Fig. \ref{fig_projection}.
Following the above guidelines, the covariance matrix estimate  $\widehat{\bM}$ is obtained as solution to the following optimization problem
\begin{equation}\label{problema_P}
{\cal{P}}\left\{
\begin{array}[c]{ll}
\underset{\bM}{\min} &
\begin{array}[c]{c}
\|\bM - \widehat{\bS}||
\end{array}
\\
\mbox{s.t.} &
\begin{array}[t]{l}
\frac{\lambda_{max}\left(\bM\right)}{\lambda_{min}\left(\bM\right)}\leq \kappa_M \\
\sigma_n^2\bI+\bR=\bM\\
\bR\succeq {\bf 0}\\
 \sigma_n^2\geq \sigma^2
\end{array}
\end{array} \right.,
\end{equation}
where $\|\cdot\|$ refers to an arbitrary but given unitary invariant matrix norm \cite{Horn} that induces a specific metric in the space of the positive semi-definite matrices over which performing the projection. Examples of norms that can be considered in the estimation process are the Frobenius, the spectral, and the Ky Fan norms. In this respect, observe that the first two instances are the most common and widely used norms in the space of Hermitian matrices corroborating the interest toward the family of unitary invariant norms.
 Remarkably, different estimators can be jointly exploited within a bank/battery of adaptive receivers. More in detail, each detector may resort to a specific norm and the presence of a prospective target is established by means of a suitable fusion logic, such as binary integration or $m$-of-$n$ detection \cite{Richards}.


Before concluding this section, it is also worth pointing out that the proposed estimators possess the consistency property as long as the secondary data vectors are statistically independent. Indeed,
\begin{eqnarray}
\|\bM - \widehat{\bM}\|&\leq&\|\bM - \widehat{\bS}\|+\|\widehat{\bS} - \widehat{\bM}\|\\
&\leq& 2\|\bM - \widehat{\bS}\|,
\end{eqnarray} 
where the first equation stems from the triangular inequality, while the second inequality follows from the  definition of $\widehat{\bM}$, i.e., it is a minimizer. Now, provided that $\widehat{\bM}$ is a measurable function of the secondary data\footnote{A sufficient condition for estimator measurability is reported in the supplementary material.}, it follows that
\begin{eqnarray}
\E[\|\bM - \widehat{\bM}\|^2]&\leq& 2\E[\|\bM - \widehat{\bS}\|^2]\\
&\leq& 2 \alpha2\E[\|\bM - \widehat{\bS}\|_F^2]\rightarrow 0,
\end{eqnarray} 
where the second inequality results from the finite dimension of the space $\mathcal{H}^N$, $\alpha$ is a specific constant linking the considered norm with the Frobenius one \cite[Corollary 5.4.5]{Horn},  and finally the convergence to zero comes from the consistency of the sample covariance estimator. Based on Chebyshev's inequality \cite{Billingsley_B}, the estimation error converges to zero also in probability which is often invoked as the classic definition of consistency.
\section{Derivation of the Structured Covariance Matrix Estimators Based on Unitary Invariant Norm Projection}\label{stimatori_matrice}
This section deals with the development of an efficient procedure to solve Problem ${\cal{P}}$ almost in closed form regardless of the considered norm. Specifically, it is proved that any sample covariance projector belongs to the class of the shrinkage estimators\footnote{A shrinkage covariance estimator $\widehat{\bM}$ is a matrix sharing the same eigenvectors as the sample covariance matrix $\bS \succeq {\bf 0}$, but transforming the eigenvalues, i.e. $\widehat{\bM}=\bU_S\, \diag([g_1(d_1,d_2,\ldots,d_N),g_2(d_1,d_2,\ldots,d_N),\ldots,\\g_N(d_1,d_2,\ldots,d_N)])\,\bU_S^{\dagger}\succeq {\bf 0}$.} and the eigenvalues estimate is obtained solving a one-dimensional convex problem whose objective function is tied up to the considered unitary norm. In general, this optimization problem can be solved in polynomial-time using convex solvers such as CVX \cite{CVX_solver}. In addition, for the two most relevant norm instances, i.e., Frobenius and spectral, very efficient algorithms are provided to tackle the associated one-dimensional optimizations.

As first step toward the solution of ${\cal{P}}$, let us observe that it is equivalent to the solvable  convex problem
\begin{equation}\label{problema_P1}
{\cal{P}}_1\left\{
\begin{array}[c]{ll}
\underset{\bX}{\min} &
\begin{array}[c]{c}
\|\bX - \bS\|
\end{array}
\\
\mbox{s.t.} &
\begin{array}[t]{l}
\bX\succeq\bI\\
\frac{\lambda_{max}\left(\bX\right)}{\lambda_{min}\left(\bX\right)}\leq \kappa_M \\
\end{array}
\end{array} \right.,
\end{equation}
where\footnote{The interested reader may refer to the supplementary material for the proof.} $\bS = \widehat{\bS}/\sigma^2$ and $\bX = \bM/\sigma^2$. Next, let us indicate with ${\bS}={\bU}_S{\bLambda}_S{\bU}^{\dag}_S$ the spectral decomposition of the normalized sample covariance matrix, where ${\bLambda}_S=\diag(\left[d_1,d_2,\cdots,d_N\right]^T)$, with $d_1\geq d_2\geq \cdots\geq d_N$ the eigenvalues of $\bS$ arranged in decreasing order, and ${\bU}_S$ is the unitary matrix whose columns contain the corresponding eigenvectors. Hence, the following lemma holds true.
\begin{lemma}\label{teorema_2_lambda}
An optimal solution to ${\cal{P}}_1$ is a shrinkage estimator $\bX^{\star}={\bU}_S{\bLambda}^\star{\bU}_S^{\dag}$ where $\bLambda^\star=\diag\left(\left[\lambda_1^\star,\lambda_2^\star, \cdots,\lambda_N^\star\right]^T\right)$ is a solution to the following optimization problem
\begin{equation}\label{problema_P2}
{\cal{P}}_2\left\{
\begin{array}[c]{ll}
\underset{\bLambda}{\min} & \|\bLambda - \bLambda_S\|\\
\mbox{s.t.} & \bLambda\succeq\bI\\
 & \frac{\lambda_{max}(\bLambda)}{\lambda_{min}(\bLambda)}\leq{\kappa_M }
\end{array}
\right.,
\end{equation}
with $\bLambda=\diag\left(\left[\lambda_1,\lambda_2, \cdots,\lambda_N\right]^T\right)$.
\end{lemma}
\proof See Appendix \ref{appendice_teorema_2_lambda}.
\endproof
To proceed further, let us introduce the auxiliary variable $u>0$ and cast ${\cal{P}}_2$ as\footnote{The interested reader may refer to the supplementary material for the proof.}
\begin{equation}\label{equation_variabile_u}
{\cal{P}}_2^{\prime}(u)\left\{
\begin{array}[c]{ll}
\underset{\bLambda,u}{\min} & \|\bLambda - \bLambda_S\|\\
\mbox{s.t.} & \lambda_i\geq1,\\
\ & u\leq{\lambda_i}\leq{\kappa_M u},\\
\ & u\geq{\frac{1}{\kappa_M }},
\end{array}
\right. \quad i=1,\ldots,N.
\end{equation}
This formulation paves the way for an efficient solution of ${\cal{P}}_2$. Indeed, for any fixed $u$ a closed form optimal matrix can be derived as shown in the following lemma.

\begin{lemma}\label{lemma_soluzione_lambda(u_segnato)}
For any $\bar{u}\geq\frac{1}{\kappa_M }$, an optimal solution $\bLambda^\star(\bar{u})$ to ${\cal{P}}_2^{\prime}(\bar{u})$,
is 
\begin{equation}
\bLambda^\star(\bar{u})=\diag(\blambda^\star(\bar{u})),
\end{equation}
where 
\begin{equation}
\begin{split}\label{espressione_di_lambda(u)}
\blambda^\star(u) = &\left[\lambda_1(u),\lambda_2(u),\ldots,\lambda_N(u)\right]^T\in\mathbb{R}^N,
\end{split}
\end{equation}
with
$$\lambda_i(u)={\min}(\kappa_M u, \max(d_i,\max(1,u))),\,\,i=1,\ldots,N.$$
\end{lemma}
\proof See Appendix \ref{appendice_lemma_soluzione_lambda(u_segnato)}.
\endproof

Leveraging on Lemmas \ref{teorema_2_lambda} and \ref{lemma_soluzione_lambda(u_segnato)}, and denoting by $g(\cdot)$ the gauge function associated with the considered
unitary invariant norm \cite{Horn}, the following fundamental result can be shown.

\begin{figure}[h!]
\begin{center}
\includegraphics[width=0.5\textwidth]{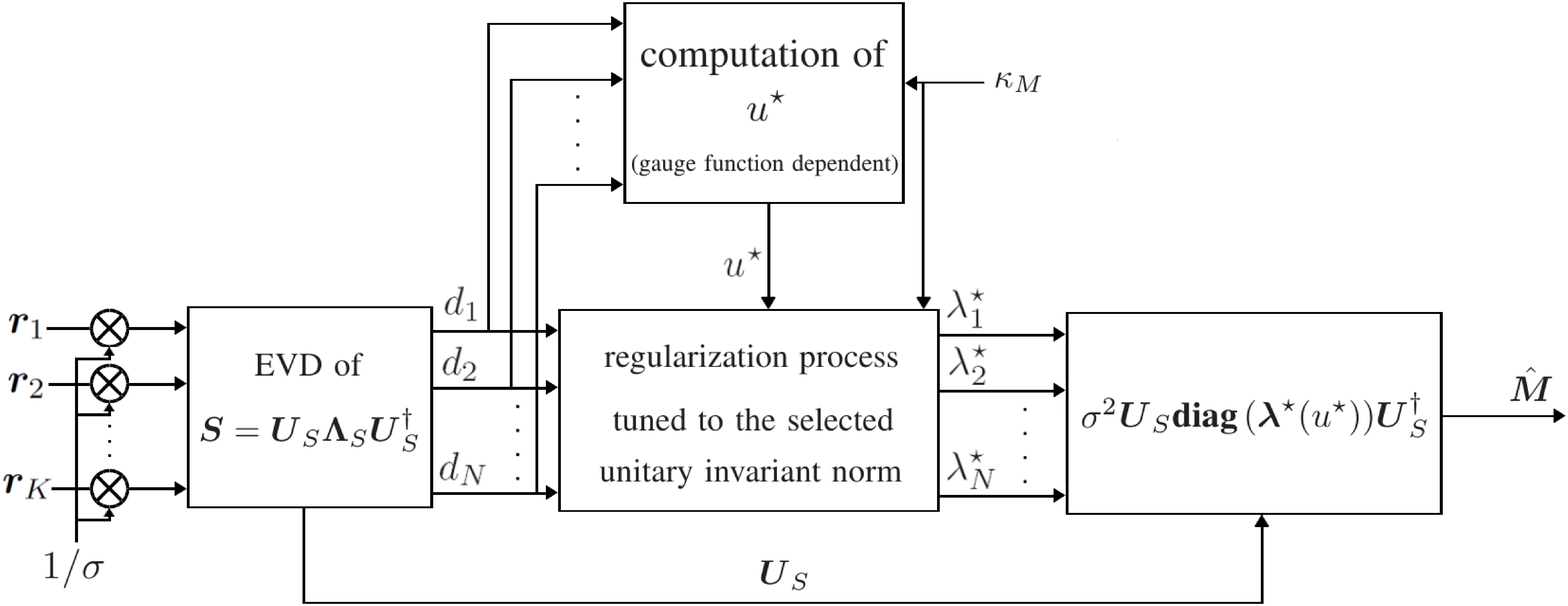}
\end{center}
\caption{Block scheme associated with the proposed estimation procedure.\label{fig_BlockScheme}}
\end{figure}

\begin{theorem}\label{espressione_finale}
Let $u^{\star}$ be the lowest optimal solution to the convex optimization problem
\begin{equation}\label{problema_P3}
{\cal{P}}_3\left\{
\begin{array}[c]{ll}
\underset{u}{\min} & g(h_1(u),h_2(u),\ldots,h_N(u))\\
\mbox{s.t.} & u\geq{\frac{1}{\kappa_M }}\\
\end{array}
\right.,
\end{equation}
where $h_i(u)=\left|\lambda_i^{\star}(u)-d_i\right|$, $i=1,\ldots,N$.  Then, an optimal solution to ${\cal{P}}_1$ is
\begin{equation}\label{risultato_stimatore}
\bX^\star={\bU}_S\,\diag(\blambda^{\star}(u^{\star})){\bU}_S^\dagger.
\end{equation}
\end{theorem}
\proof See Appendix \ref{espressione_finale_prrof}.
\endproof
According to Theorem \ref{espressione_finale},  a unique solution to ${\cal{P}}_1$ can be constructed in almost closed form. Consequently, this new class of estimators effectively performs specific unitary invariant norm-based projections. It is also worth pointing out that the functional dependence over the selected  norm is concentrated in the optimal value of the auxiliary $u$ that accounts for the corresponding gauge function. {In Fig. \ref{fig_BlockScheme}, a schematic illustration of the steps involved in the proposed procedure for the computation of $\widehat{\bM}$ is reported. As already highlighted, $\widehat{\bM}$ is a shrinkage estimator which
regularizes the sample covariance matrix according to the specific unitary invariant norm and explicitly accounting for a condition number constraint so as to provide a well-conditioned structured estimate.}

Before continue further, it is worth observing that the functions $h_i(u)$, $i=1,\ldots, N$, involved in ${\cal{P}}_3$ can be expressed in closed form\footnote{The interested reader may refer to the supplementary material for the proof.}. This is a useful result in general and more important it lays the ground for the derivation of specialized procedures to handle ${\cal{P}}_3$ assuming either the Frobenius or the spectral norm (see Subsections \ref{sezione_frobenius} and \ref{sezione_spettrale}). Precisely, if $d_i>1$
\begin{equation}\label{appendice_h(u)caso_d_i>1}
h_i(u) = \left\{
\begin{array}[c]{ll}
d_i - \kappa_M u & \mbox{if}\quad \frac{1}{\kappa_M }\leq u < \frac{d_i}{\kappa_M }\\
0 & \mbox{if}\quad \frac{d_i}{\kappa_M }\leq u < d_i\\
u - d_i & \mbox{if}\quad u \geq d_i\\
\end{array}
\right.,
\end{equation}
otherwise $d_i \leq 1$ and
\begin{equation}\label{appendice_h(u)caso_d_i<1} 
h_i(u) = \left\{
\begin{array}[c]{ll}
1 - d_i & \mbox{if}\quad \frac{1}{\kappa_M }\leq u < 1\\
u - d_i & \mbox{if}\quad u\geq 1\\
\end{array}
\right..
\end{equation}

\subsection{Frobenius Norm}\label{sezione_frobenius}
The gauge function associated with the Frobenius norm is given by 
$$g(h_1(u),h_2(u),\ldots,h_N(u))=\sqrt{\displaystyle{\sum_{i=1}^N}{(h_i(u)})^2}$$
and the problem to solve boils down to 
\begin{equation}\label{frob_obiettivo}
{\cal{P}}_4\left\{
\begin{array}[c]{ll}
\underset{u}{\min} & G_1(u)\\
\mbox{s.t.} & u \geq \frac{1}{\kappa_M }\\
\end{array}
\right.,
\end{equation}
where $G_1(u)=\displaystyle{\sum_{i=1}^N}{(h_i(u))^2}$. To proceed further, let us observe that the optimal solution to problem ${\cal{P}}_4$ is $u^{\star} = \frac{\max\{1,d_1\}}{\kappa_M }$ if $d_1\leq \kappa_M $, otherwise, the optimal solution to ${\cal{P}}_4$ lies within the interval\footnote{The interested reader may refer to the supplementary material for the proof.} $[1, d_1]$. In this last case, as shown in supplementary material, $G_1(u)$  is a convex function with a continuous derivative function within the interval of interest $\left[1,d_1\right]$. 
Hence, denoting by $\bar{N}$ the number of sample eigenvalues $d_i$ greater than $1$, i.e. $d_i>1$, $i=1,\ldots,\bar{N}$, and defining the vector
$$
\bv=\left[d_1,d_2,\ldots,d_{\bar{N}},1\right]^T\in\mathbb{R}^{\bar{N}+1},
$$
the following theorem holds true.
\begin{theorem}\label{teorema_soluzione_d1_magg_Kmax}
Assuming $d_1 > \kappa_M>1 $, the optimal solution $u^{\star}$ to ${\cal{P}}_4$ is
\begin{enumerate}
\item $u^{\star} = 1$, if $\frac{d G_1(u)}{d u}\Big|_{u=1} \geq 0$;
\item $u^{\star} = d_1$, if $\frac{d G_1(u)}{d u}\Big|_{u=d_1}\leq 0$;
\item $u^{\star}=\frac{d_1}{\kappa_M}$, if $\frac{d_1}{\kappa_M} \leq d_N$;
\item if $1)$, $2)$, and $3)$ are not satisfied, $u^{\star}$ is the optimal solution if and only if
\begin{equation}\label{u_star_con_algoritmo}
u^{\star} = \frac{\displaystyle{\sum_{i=1}^{\beta}} \kappa_M d_i + \displaystyle{\sum_{i=\alpha}^{N}} d_i}{N-\alpha +1 +{\beta}\kappa_M ^2},
\end{equation}
with $\alpha \in \{1,2,\ldots,\bar{N},\bar{N}+1\}$ the smallest index such that $v_\alpha < u^{\star}$, and $\beta \in \{1,2,\ldots,\bar{N},\bar{N}+1\}$ the largest index such that $\frac{v_\beta}{\kappa_M } > u^{\star}$.
\end{enumerate}
\end{theorem}
\proof See Appendix \ref{appendice_teorema_soluzione_d1_magg_Kmax}.
\endproof
Theorem \ref{teorema_soluzione_d1_magg_Kmax} provides the guidelines to find $u^{\star}$. Indeed, the selection of the integers $\alpha$ and $\beta$ such that
\begin{equation}\label{u_con_alfa_beta}
u_{\alpha,\beta} = \frac{\displaystyle{\sum_{i=1}^{\beta}} \kappa_M  d_i + \displaystyle{\sum_{i=\alpha}^{N}}d_i}{N-\alpha +1 + \beta \kappa_M ^2},
\end{equation}
with
\begin{equation}\label{test1}
v_\alpha < u_{\alpha,\beta} \leq v_{\alpha-1} \quad \mbox{and} \quad
\frac{v_{\beta+1}}{\kappa_M }\leq u_{\alpha,\beta} < \frac{v_{\beta}}{\kappa_M },
\end{equation}
is required. To this end, an efficient procedure is now described. Precisely, the strategy consists in iteratively verifying the conditions \eqref{u_con_alfa_beta} and \eqref{test1} once the values of $\alpha$ and $\beta$ have been efficiently fixed. In this respect, notice that, if the intersection of the intervals \eqref{test1} is empty, then $u_{\alpha,\beta}$ cannot be the optimal solution. On the contrary, the intersection is given by one of the following sub-intervals $\Big]v_\alpha, v_{\alpha-1}\Big], \left]v_\alpha, \frac{v_{\beta}}{\kappa_M }\right[, \left[\frac{v_{\beta+1}}{\kappa_M }, v_{\alpha-1}\right], \left[\frac{v_{\beta+1}}{\kappa_M }, \frac{v_{\beta}}{\kappa_M }\right[$, and the optimal point must belong to one of them. According to this line of reasoning, the procedure follows these simple steps:
\begin{enumerate}
\item Set $\beta=1$, $\alpha=2$, and increase $\alpha$ until $v_{\alpha} \geq \frac{v_{\beta}}{\kappa_M }$.
\item Compute $u_{\alpha,\beta}$. If $u_{\alpha,\beta}$ belongs to the current intersection, let $u^{\star} = u_{\alpha,\beta}$ and exit; otherwise go to step $3)$.
\item If $ \frac{v_{\beta+1}}{\kappa_M }<v_{\alpha} $, increase $\alpha$ and go to step $2)$, otherwise increase $\beta$ and go to step $2)$.
\end{enumerate}

It is worth pointing out that this algorithm provides the optimal solution to problem ${\cal{P}}_4$ with a linear computational complexity with the number of the sample covariance matrix eigenvalues greater than $1$.

\subsection{Spectral Norm}\label{sezione_spettrale}
In this case, the gauge function is 
$$g(h_1(u),h_2(u),\ldots,h_N(u))=\displaystyle{\max_{i=1,\ldots,N}}\left\{{h_i(u)}\right\}$$
and ${\cal{P}}_3$ can be specialized as

\begin{equation}\label{frb_obiettivo}
\bar{{\cal{P}}}_3\left\{
\begin{array}[c]{ll}
\underset{u}{\min} & G_2(u)\\
\mbox{s.t.} & u \geq \frac{1}{\kappa_M }\\
\end{array}
\right.,
\end{equation}
with $G_2(u)=\underset{i}{\max}\left\{h_i(u)\right\}$.

The following proposition provides  an efficient procedure to solve  $\bar{{\cal{P}}}_3$.
\begin{theorem}\label{sp_funzione_G(u)}
Let $u^{\star}$ the lowest optimal solution to  $\bar{{\cal{P}}}_3$. Then 
\begin{enumerate}
\item  $u^{\star}=\frac{1}{\kappa_M}$, if $d_1\leq 1$.
\item $u^{\star}=\max\left\{\frac{d_1+d_N-1}{\kappa_M},\frac{1}{\kappa_M}\right\}$, if $1 < d_1\leq \kappa_M $ and $d_N \leq 1$.
\item $u^{\star} = \frac{d_1}{\kappa_M}$, if $1 < d_1\leq \kappa_M $ and $d_N > 1$. In this case, the covariance estimate reduces to $\bS$.
\item If $d_1>\kappa_M $ and $d_N\leq 1$, then
\begin{itemize}
	\item $u^{\star}=\max\left\{\eta_1,\frac{1}{\kappa_M}\right\}$, if $\eta_1=\frac{d_1+d_N-1}{\kappa_M } \leq 1$;
	\item otherwise, $u^{\star}=\frac{d_1+d_N}{1+\kappa_M }>1$.
\end{itemize}
\item If $d_1>\kappa_M $ and $d_N>1$, then
	\begin{itemize}
		\item  $u^{\star} = \frac{d_1+d_N}{1+\kappa_M }$. if $d_N \leq \frac{d_1}{\kappa_M }$;
		\item otherwise, $u^{\star} = \frac{d_1}{\kappa_M }$. Besides, in this case, the estimate coincides with $\bS$.
	\end{itemize}
\end{enumerate}
\end{theorem}
\proof See Appendix \ref{appendice_sp_funzione_G(u)}.
\endproof

Based on Theorem \ref{sp_funzione_G(u)}, the optimal solution to $\bar{{\cal{P}}}_3$ is substantially available in closed form. Indeed,  just  the comparison between some linear functions of the highest and the lowest sample covariance eigenvalues with some fixed thresholds is required. It is also worth pointing out that, $u^\star$ is a continuous function of $d_1,\ldots,d_N$ implying that $\bX^\star$ is a continuous function of $\bS$.
\section{Performance Analysis}\label{risultati}

This section is devoted to the analysis of the proposed covariance estimators. The average SINR is adopted as performance metric and some counterparts available in the open literature are considered for comparison purposes. Two typical radar signal processing scenarios are studied: the former focuses on spatial processing with wideband jammers impairing the received data, the latter considers Doppler processing with the interfering returns originated by clutter.
More formally, the average SINR (over $MC$ i.i.d. realizations\footnote{In the numerical results the number of Monte Carlo trials is set to $MC=500$.} of $K$ secondary data) is given by
\begin{equation}\label{sinr}
\mbox{SINR}_{av} = \frac{1}{MC}\sum_{i=1}^{MC}\frac{|\widehat{\bw}_i^{\dag}\bs(x)|^2}{\left(\widehat{\bw}_i^{\dag}\bM \widehat{\bw}_i\right)},
\end{equation}
where $\bs(x)$ is the $N$-dimensional target steering vector whose expression depends on:
\begin{itemize}
\item the considered processing scenario (spatial/temporal);
\item the radar configuration (array type, Pulse Repetition Time (PRT), etc.);
\item the target state $x$ (angle of arrival $\theta$/normalized Doppler frequency $\nu$).
\end{itemize} 
Moreover, $\widehat{\bw}_i=\widehat{\bM}_i^{-1}\bs(x)$ is the adaptive estimate of the optimal weight vector, where $\widehat{\bM}_i$ is the data-dependent estimate of $\bM$ at the $i$-th run. 

In the following analysis, it is assumed $\sigma^2=0$ dB and $\kappa_M =\lambda_{max}(\bM)/\lambda_{min}(\bM)$.
Furthermore, for each case study two different values of the actual white noise power level are considered, i.e., $\sigma_a^2=0$ and $\sigma_a^2=10$ dB, so as to account for both a matched and a mismatched scenario. 

\subsection{Spatial Processing}\label{spatial_processing}

A radar system equipped with a uniform linear array of $N=8$ elements (with a spacing between the antennas of $d=\lambda_0/2$ where $\lambda_0$ is the radar operating wavelength) pointing in the boresight direction is considered. The interference covariance matrix is given by  $\bM = \bM_s + \sigma_a^2\bI$ \cite{Gerlach} where $\sigma_a^2$ is the actual power level of the white disturbance term, whereas $\bM_s$ is the covariance matrix associated to $J$ (possibly wideband) jammers. Specifically, $\forall\left(n,m\right)\in\left\{1,\ldots,N\right\}^2$, 
\begin{equation}\label{matrice_cov2}
\begin{split}
\bM_s\left(n,m\right) = \displaystyle{\sum_{i=1}^J} &\sigma_i^2\mbox{sinc}\left[0.5B_f(n-m)\phi_i\right]e^{j(n-m)\phi_i},\\ 
\end{split}
\end{equation}
with $B_f=B/f_0$  the fractional bandwidth, $B$ the instantaneous bandwidth of the desired signal (coinciding with the jammer's bandwidth), $f_0 = c/\lambda_0$, $\sigma_i^2$ the power associated with the $i$-th jammer, and $\phi_i$ the jammer phase angle with respect to the antenna phase center. Precisely, $\phi_i=2\pi d(\sin\theta_i)/\lambda_0$, with $\theta_i$ the angle off-boresight of the jammer. Finally, according to the specified system model, the steering vector \eqref{sinr} reduces to $$\bs(\theta)=\left[1, \exp\left(j\pi\sin(\theta)\right), \ldots, \exp\left(j\pi\sin(\theta)(N-1)\right)\right]^T.$$

\begin{figure*}[ht!]
\begin{center}
\subfigure[$\sigma_a^2=0$ dB, $K=4$]{\includegraphics[width=0.49\textwidth]{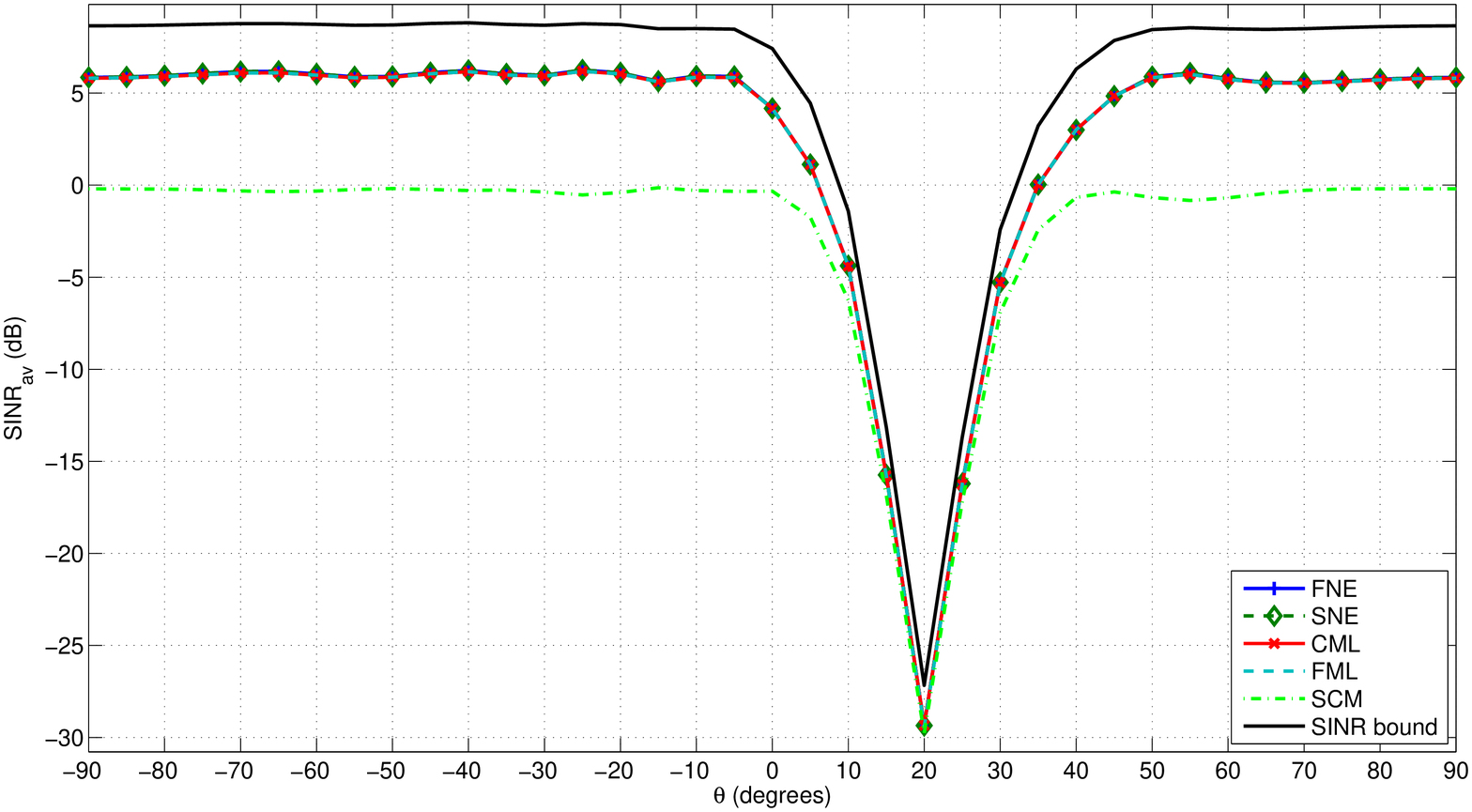}}
\subfigure[$\sigma_a^2=10$ dB, $K=4$]{\includegraphics[width=0.49\textwidth]{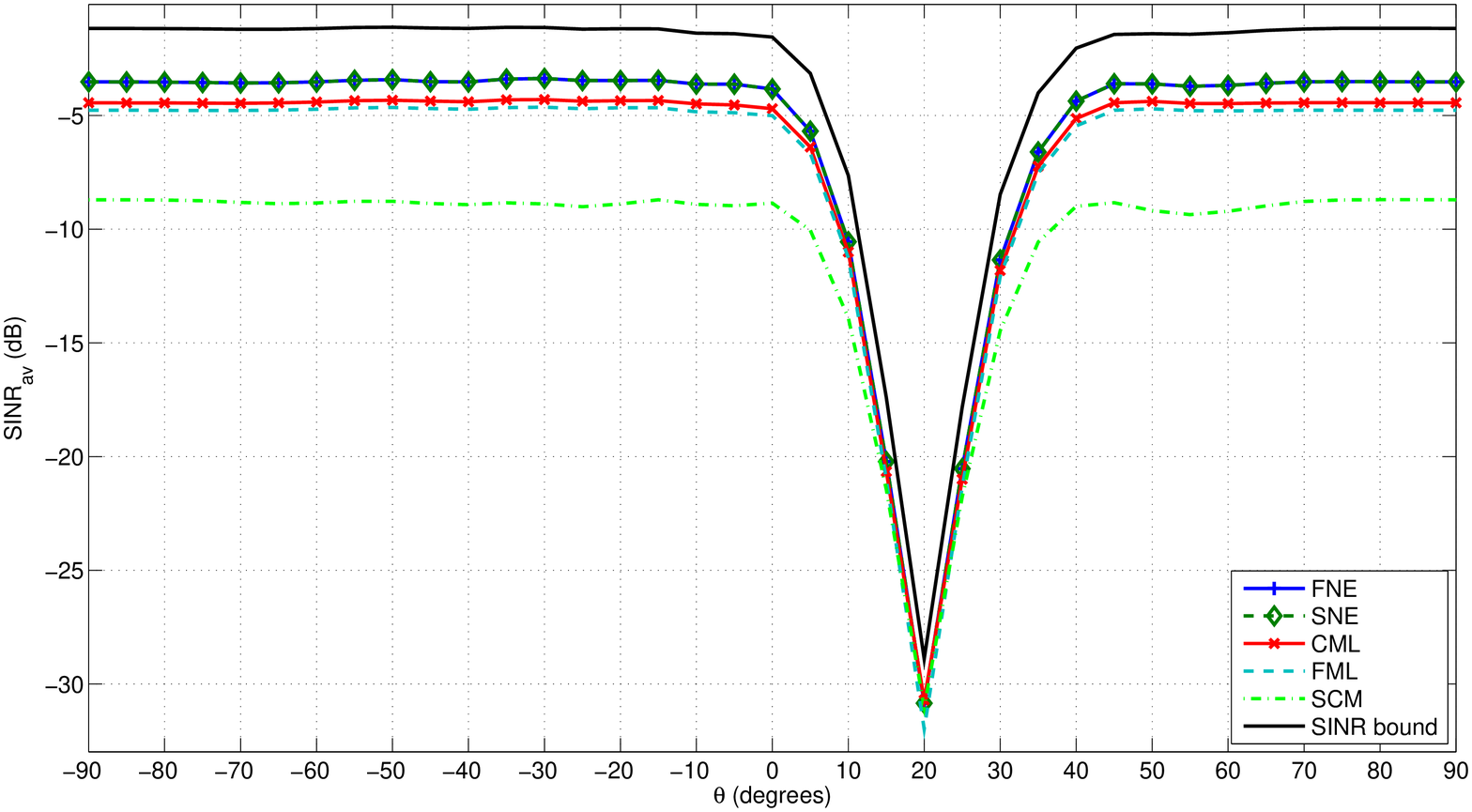}}
\subfigure[$\sigma_a^2=0$ dB, $K=8$]{\includegraphics[width=0.49\textwidth]{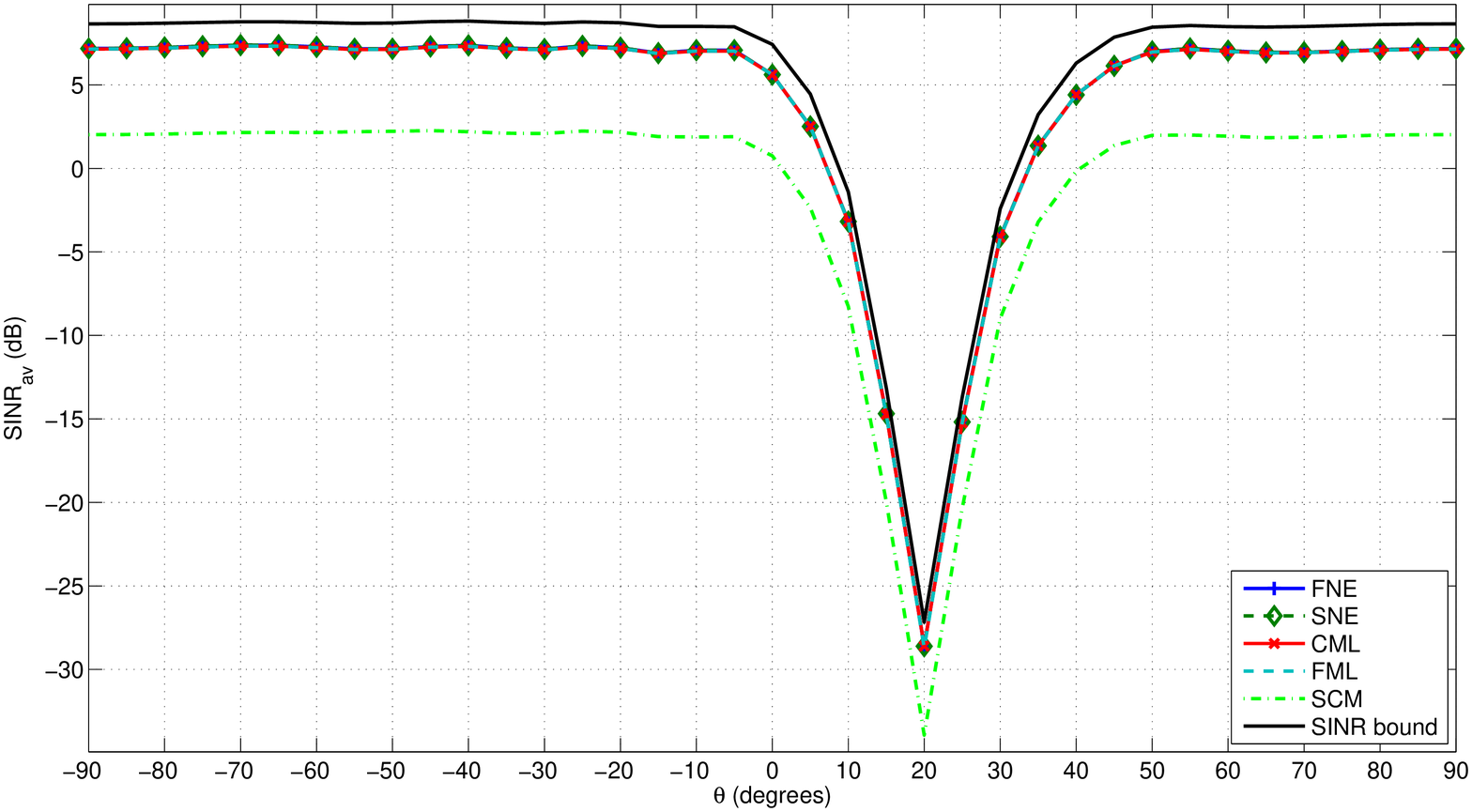}}
\subfigure[$\sigma_a^2=10$ dB, $K=8$]{\includegraphics[width=0.49\textwidth]{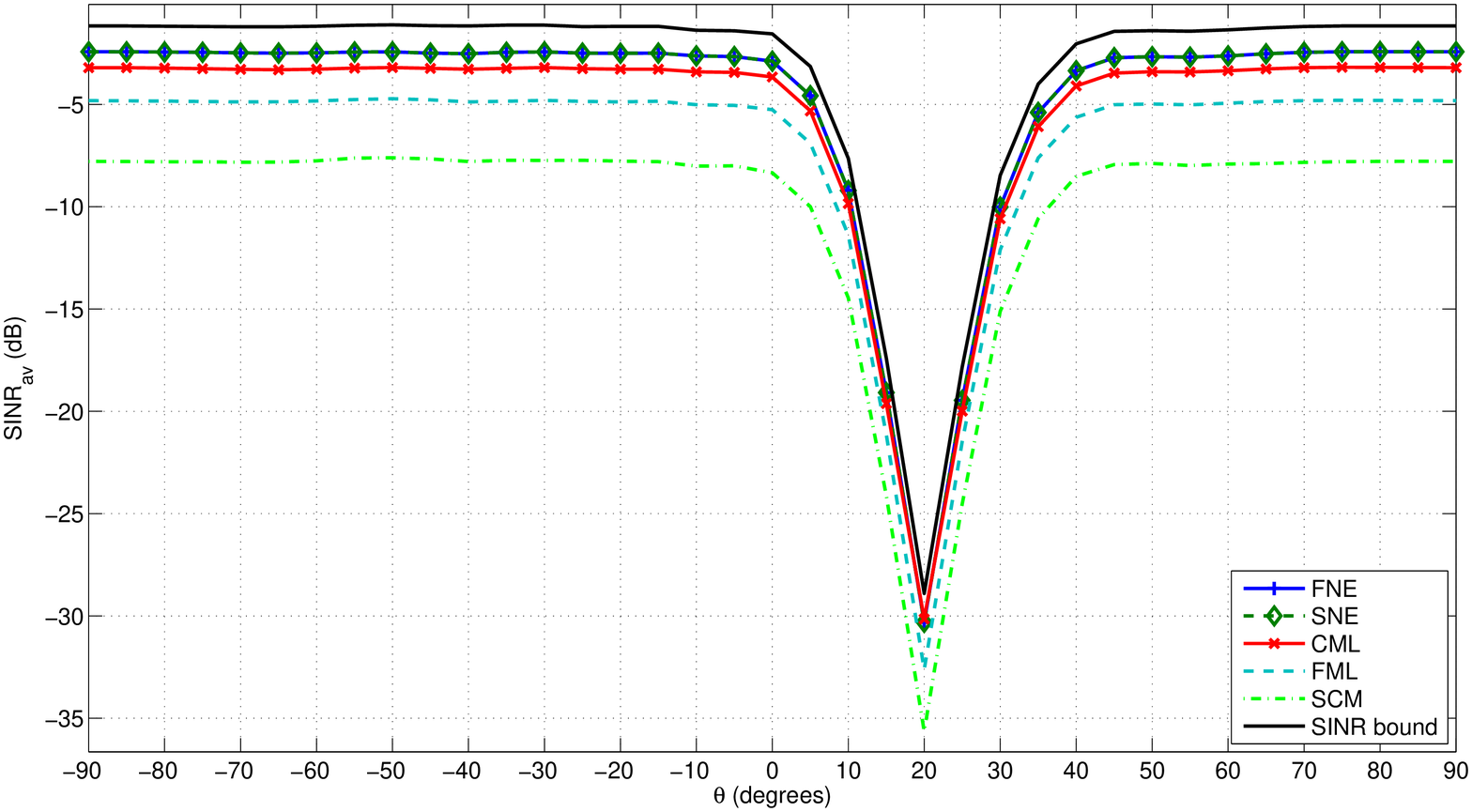}}
\subfigure[$\sigma_a^2=0$ dB, $K=16$]{\includegraphics[width=0.49\textwidth]{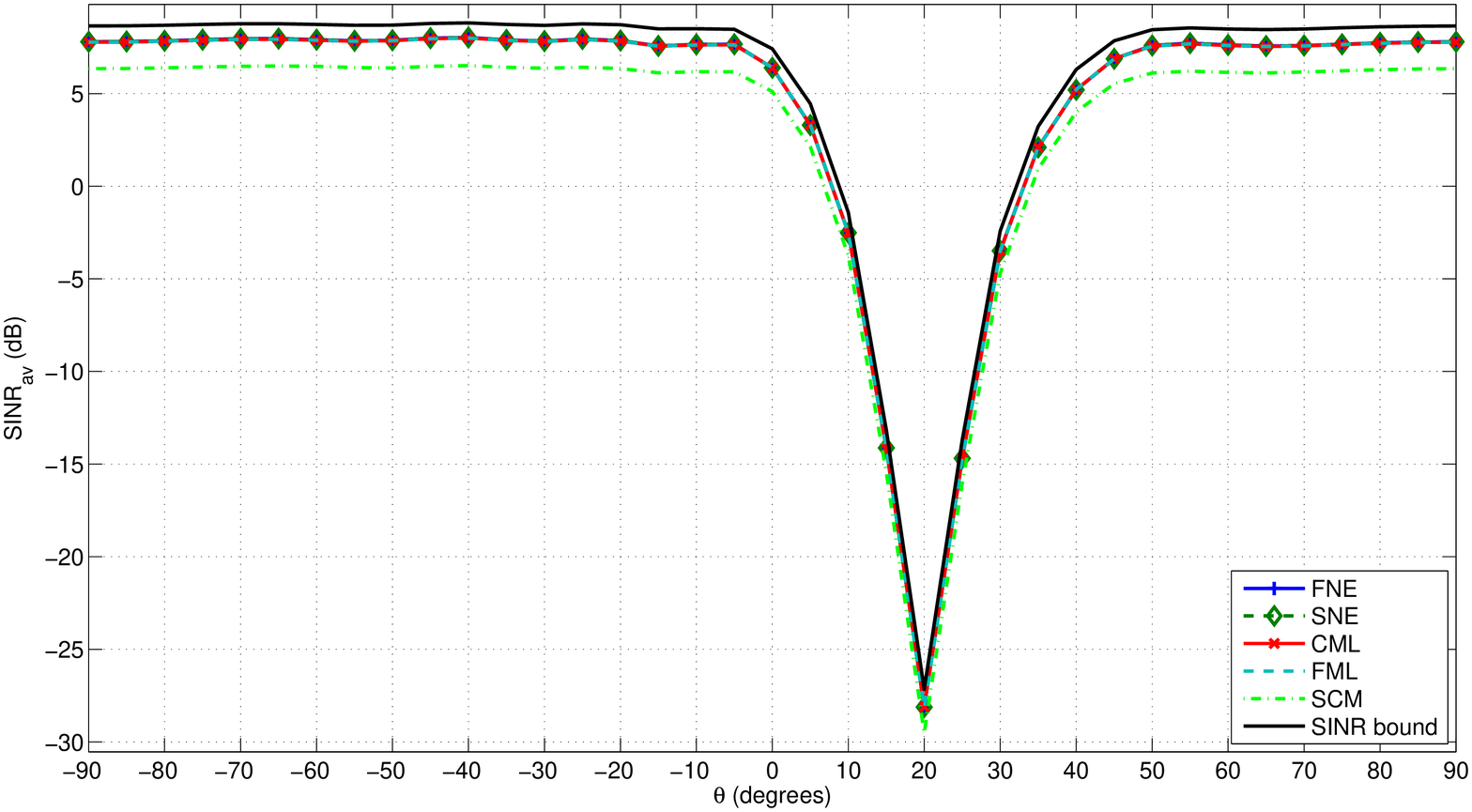}}
\subfigure[$\sigma_a^2=10$ dB, $K=16$]{\includegraphics[width=0.49\textwidth]{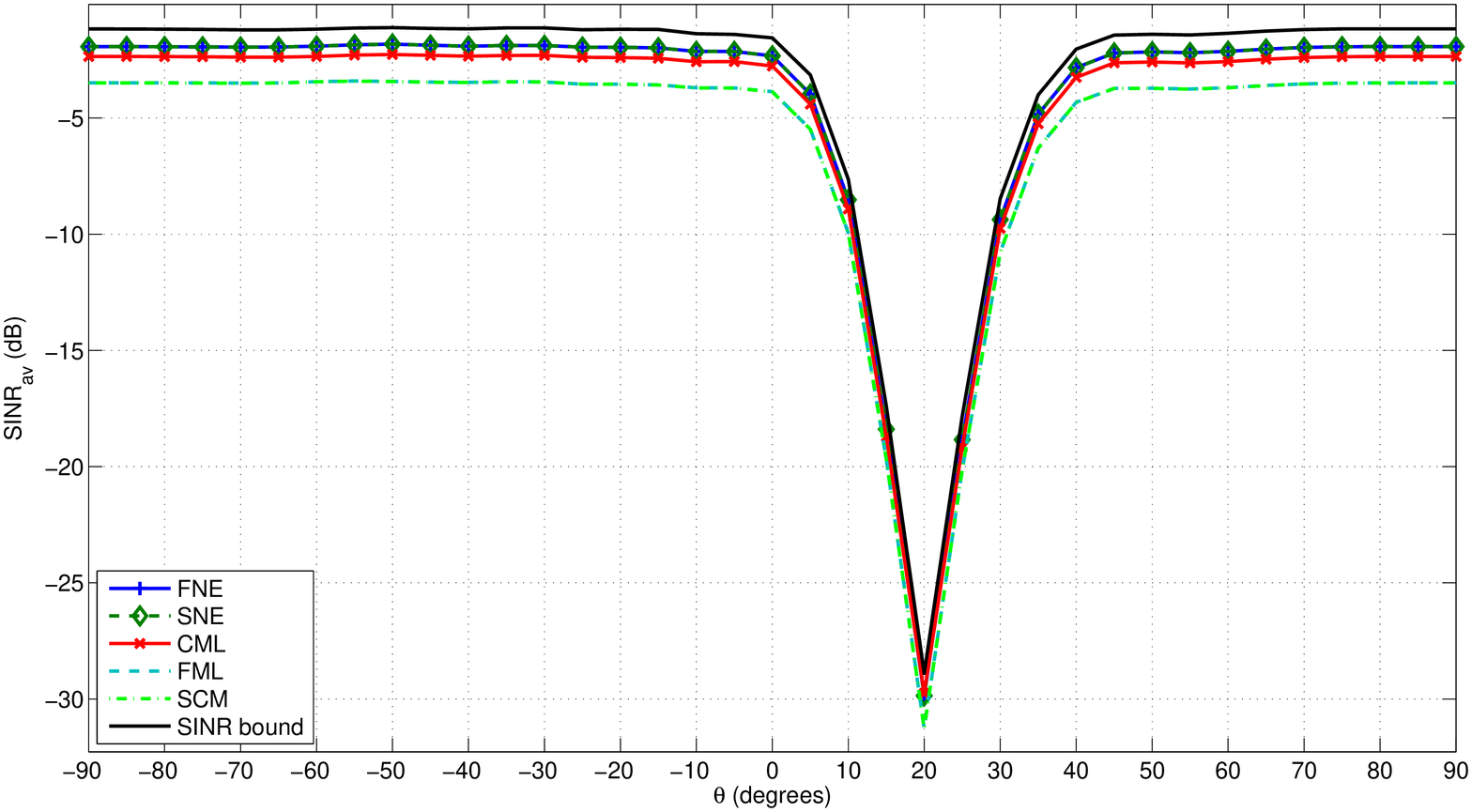}}
\end{center}
\caption{Spatial processing in the presence of Gaussian secondary data for different numbers of training data. $\mbox{SINR}_{av}$ versus $\theta$ (blue curve $+$-marked FNE, green dashed curve SNE, red $\times$-marked curve CML, cyan dashed curve FML, green dot-dashed curve SCM, and the black curve ideal SINR bound). A wideband jammer with $B_{f_1}=0.3$, $\sigma^2_1=30$ dB and $\theta_1=20$ deg is present. Subplots (a), (c), and (e) refer to the case $\sigma_a^2=0$ dB, whereas subplots (b), (d), and (f) to $\sigma_a^2=10$ dB.\label{fig_gauss_spatial}}
\end{figure*}

As case study, it is considered a wideband jammer with a fractional bandwidth $B_{f_1} = 0.3$, a power $\sigma_1^2 = 30$ dB, and a direction of arrival $\theta_1 = 20$ deg that impinges of the radar receive array. In Fig. \ref{fig_gauss_spatial},  the average SINR is plotted versus the Direction Of Arrival (DOA) $\theta$  for both the Frobenius Norm based Estimator (FNE) and the Spectral Norm based Estimator (SNE). Therein, the secondary data are modeled as i.i.d., zero-mean, circularly symmetric Gaussian random vectors. For comparison purposes, the SINR behavior associated with the Constrained ML estimator (CML) \cite{TSP}, the Fast Maximum Likelihood estimator (FML)  \cite{Gerlach}, and the classic Sample Covariance Matrix (SCM)\footnote{Notice that, when $K<N$ the pseudo inverse of the sample matrix is utilized in place of its inverse \cite{Reed}.} is displayed too. Besides, the SINR upper bound $\bs^{\dag}\bM^{-1}\bs$ is reported as benchmark. Figs. \ref{fig_gauss_spatial}(a), \ref{fig_gauss_spatial}(c), \ref{fig_gauss_spatial}(e) assume a matched condition, i.e. $\sigma_a^2=\sigma^2=0$ dB, with $K=4$, $K=8$, and $K=16$ secondary data, respectively. Figs. \ref{fig_gauss_spatial}(b), \ref{fig_gauss_spatial}(d), \ref{fig_gauss_spatial}(e) account for a mismatched  situation, i.e. $\sigma_a^2=10$ dB and $\sigma^2=0$ dB. 

The results show that FNE and SNE substantially exhibit the same performance in terms of average SINR regardless of the considered DOA and scenario. In the matched cases the curves of the new devised estimators almost overlap with those of FML and CML  (the maximum gain of FNE and SNE over CML and FML is 0.08 dB) and significantly dominate the SCM performance. Besides, FNE and SNE outperform all the counterparts in the mismatched scenario with gains (with respect to CML, FML, and SCM) of: $0.93$ dB, $1.26$ dB, and $5.65$ dB for $K=4$; $0.80$ dB, $2.40$ dB, and $5.45$ dB for $K=8$;  $0.45$ dB, $1.60$ dB, and $1.60$ dB for $K=16$. As expected, increasing the sample support size the gain reduces since all the curves tend to approach the SINR upper bound due to the consistency of the estimators.

\begin{figure*}[ht!]
\begin{center}
\subfigure[$\sigma_a^2=0$ dB, $K=4$]{\includegraphics[width=0.49\textwidth]{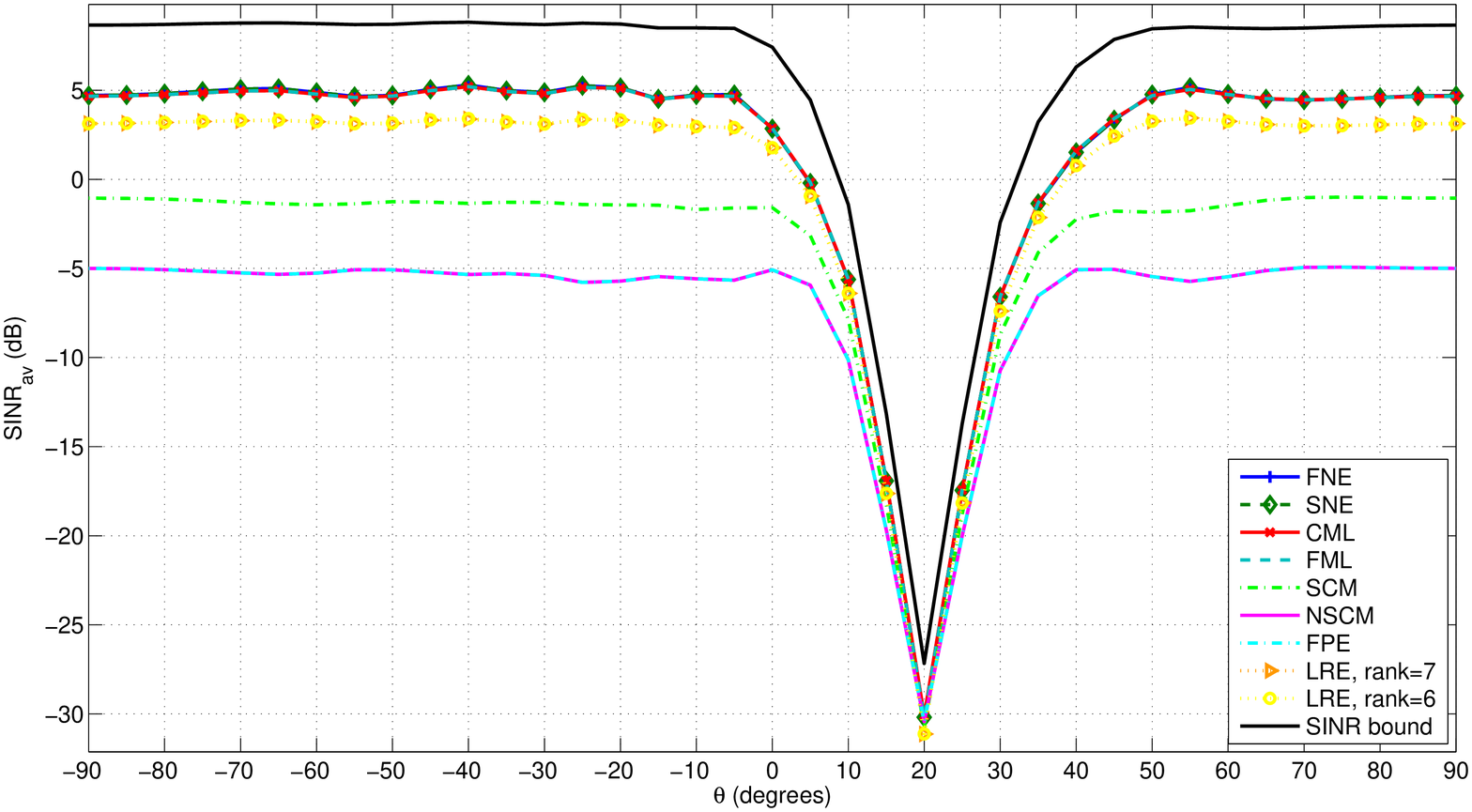}}
\subfigure[$\sigma_a^2=10$ dB, $K=4$]{\includegraphics[width=0.49\textwidth]{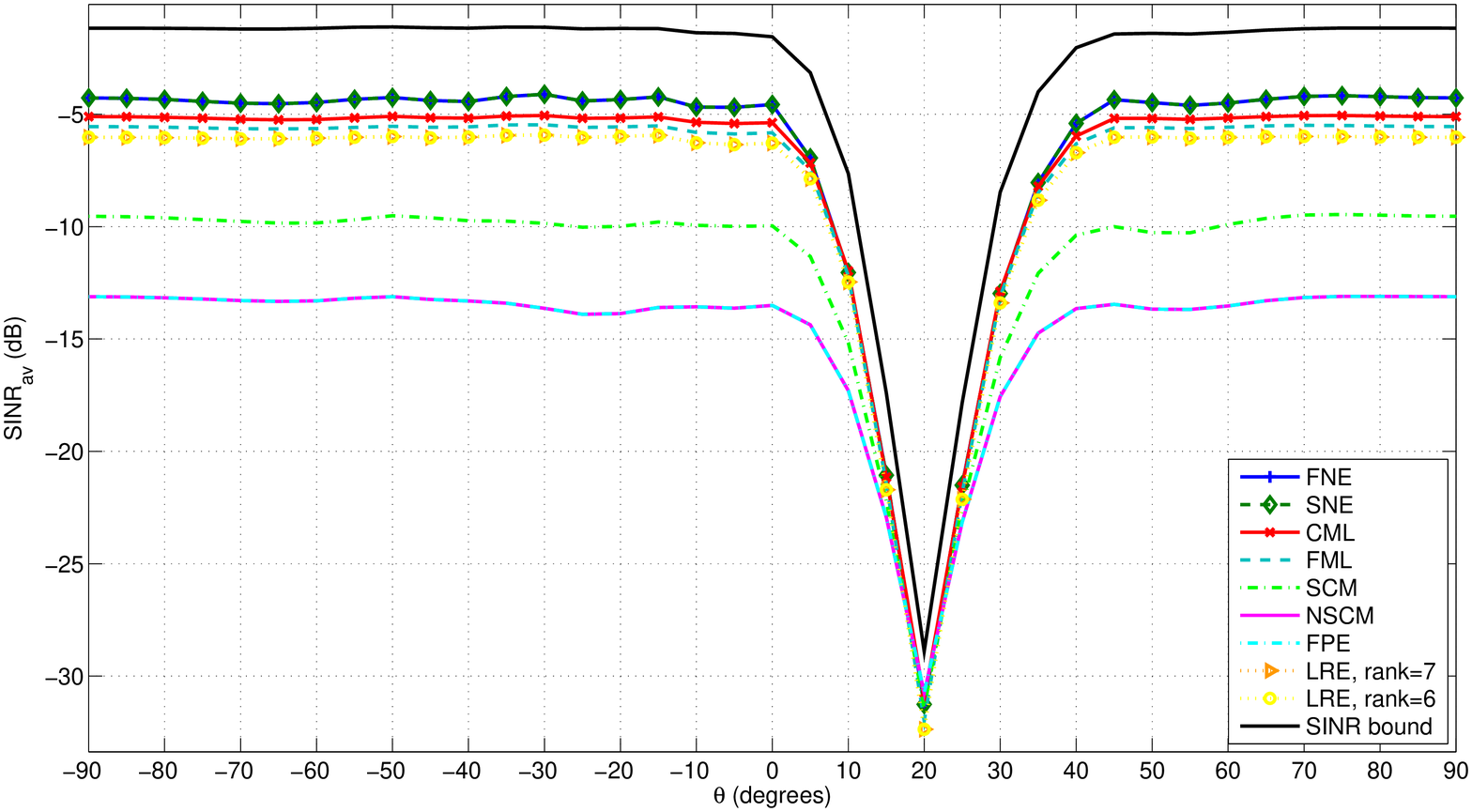}}
\subfigure[$\sigma_a^2=0$ dB, $K=8$]{\includegraphics[width=0.49\textwidth]{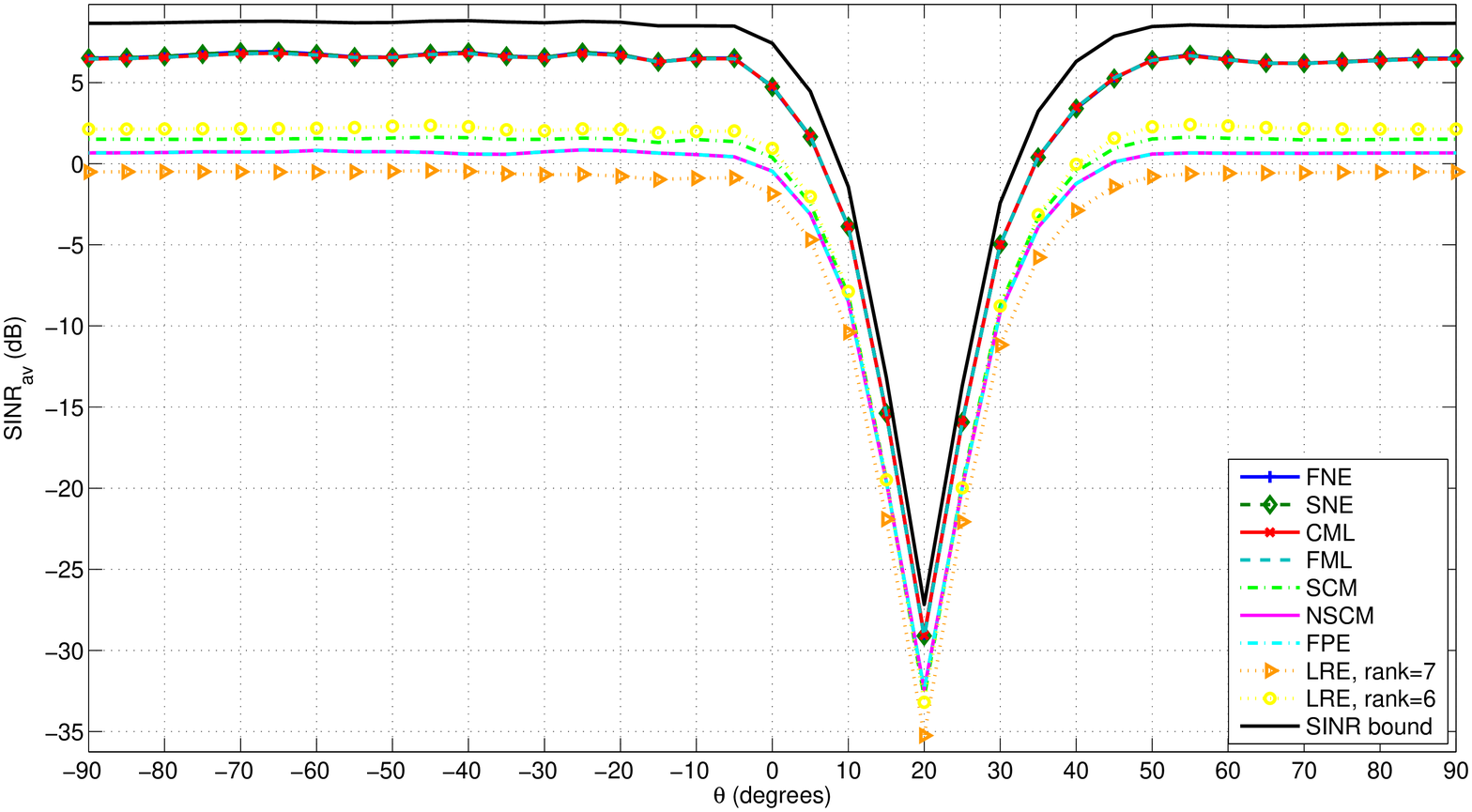}}
\subfigure[$\sigma_a^2=10$ dB, $K=8$]{\includegraphics[width=0.49\textwidth]{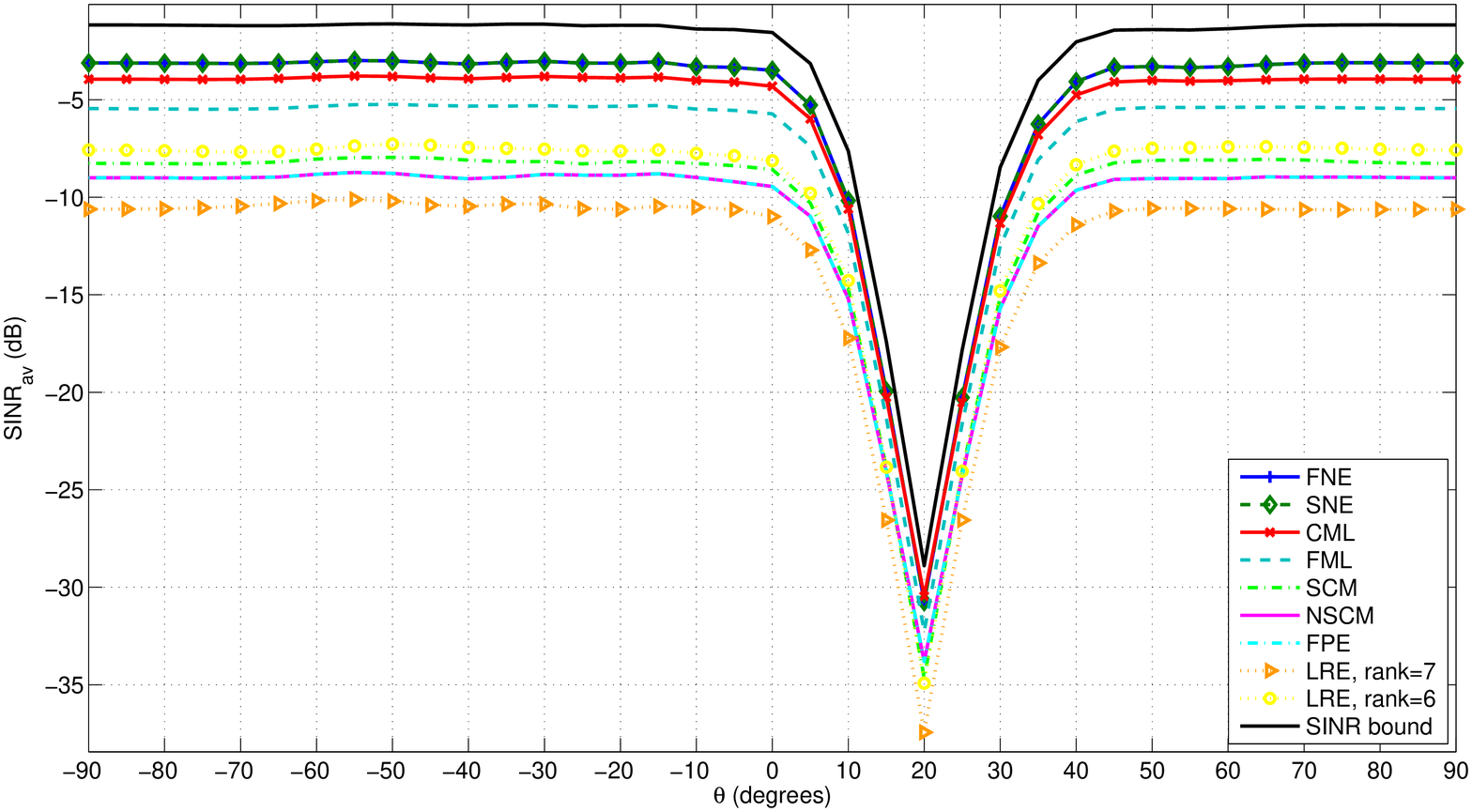}}
\subfigure[$\sigma_a^2=0$ dB, $K=16$]{\includegraphics[width=0.49\textwidth]{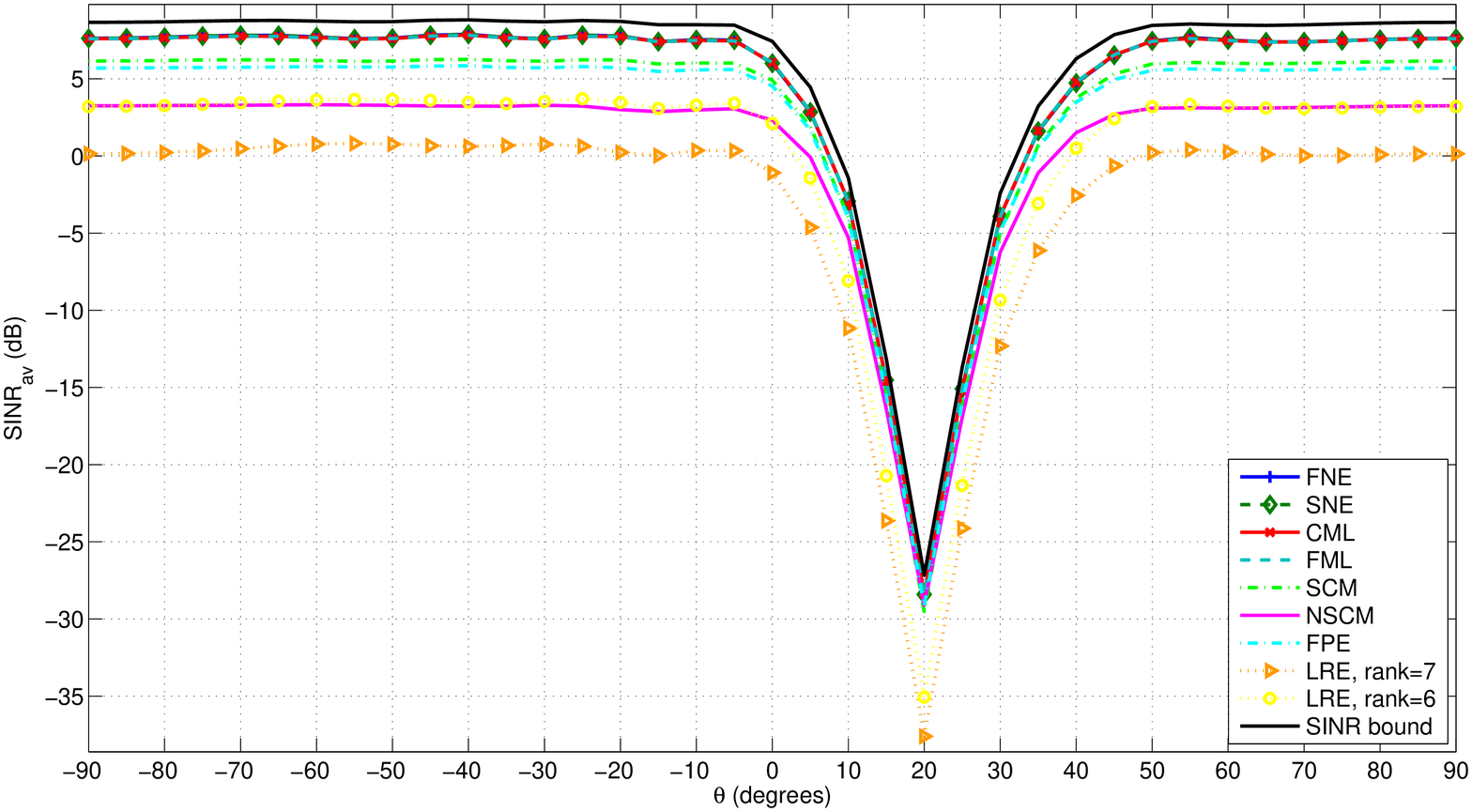}}
\subfigure[$\sigma_a^2=10$ dB, $K=16$]{\includegraphics[width=0.49\textwidth]{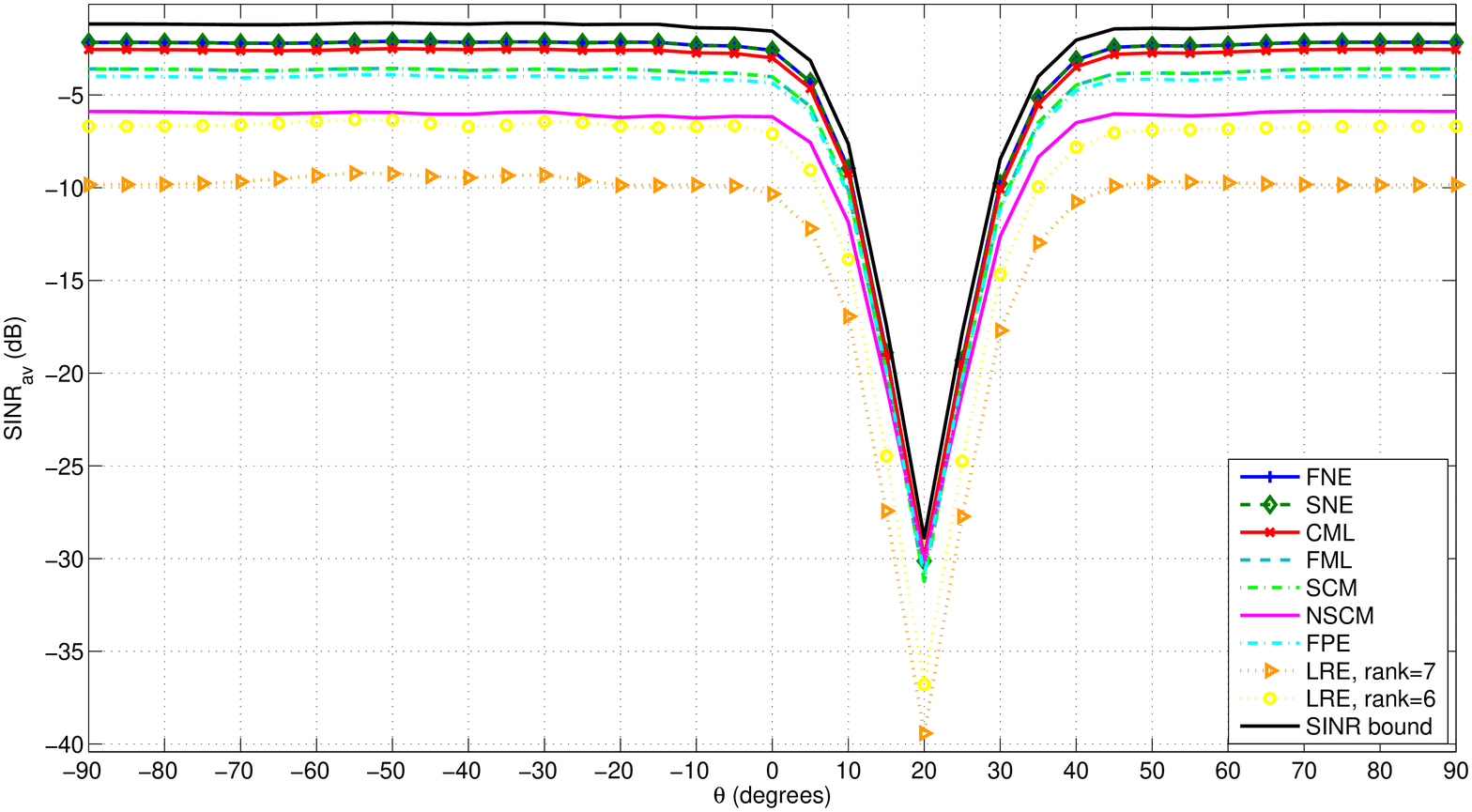}}
\end{center}
\caption{Spatial processing in the presence of mixed Gaussian and compound Gaussian data. $\mbox{SINR}_{av}$ versus $\theta$ (blue curve $+$-marked  FNE, green dashed curve SNE, red $\times$-marked curve CML, cyan dashed curve FML, green dot-dashed curve SCM, magenta curve NSCM, cyan dot-dashed curve FPE, orange dotted curve $\triangleright$-marked LRE-7 , yellow dotted curve $\circ$-marked LRE-6, and black curve ideal SINR bound). A wideband jammer with $B_{f_1}=0.3$, $\sigma^2_1=30$ dB and $\theta_1=20$ deg is present. Subplots (a), (c), and (e) refer to the case $\sigma_a^2=0$ dB, whereas subplots (b), (d), and (f) to $\sigma_a^2=10$ dB.\label{fig_compound_spatial}}
\end{figure*}

In Fig. \ref{fig_compound_spatial}, the same spatial processing scenario as in Fig. \ref{fig_gauss_spatial} is analyzed, but for a different secondary data statistical distribution that is no longer Gaussian. Specifically, 
\begin{equation}\nonumber
\br_i = \bn_i + \sqrt{\tau_i}\bx_i\, \quad i=1,\ldots,K,
\end{equation}
where $\bn_i\sim \mathcal{CN}(0,\sigma_a^2 \bI)$, $\bx_i \sim \mathcal{CN}(0,\bM_s)$, and $\tau_i \sim \Gamma(1/\mu_{\tau},\mu_{\tau})$ ($\mu_{\tau}=2$), $i=1,\ldots,K$,  are statistically independent random variables/vectors.  Otherwise stated,  a compound  Gaussian jamming is now accounted for. 
As in Fig. \ref{fig_gauss_spatial}, the average SINR versus $\theta$ is reported for three different sample support sizes, i.e., $K=4$, $K=8$, and $K=16$. Moreover, the subplots on the left refer to the matched scenario ($\sigma_a^2=\sigma^2=0$ dB)  whereas the plot on the right address the mismatched case ($\sigma_a^2=10$ dB and $\sigma^2=0$ dB).

For comparison purposes, other than the FML, CML, and SCM estimators, three additional strategies designed to operate in compound Gaussian clutter are considered. Specifically, 
\begin{itemize}
\item the Normalized Sample Covariance Matrix estimator (NSCM), \cite{gini2000}
\begin{equation}\nonumber
\widehat{\bM}_{\text{NSCM}} = \frac{N}{K}\sum_{i=1}^K \frac{\br_i\br_i^{\dag}}{\br_i^{\dag}\br_i};
\end{equation}
\item the Fixed-Point Estimator (FPE) \cite{Pascal2008,Chitour2008,ConteDeMaioRicci} $\widehat{\bM}_{\text{FPE}}$, that is obtained iteratively solving a fixed point equation;
\item the Low Rank clutter Estimator (LRE), \cite{SunPalomar}, addressing a mixed Gaussian/compound Gaussian disturbance model. {The covariance estimate can be computed as
\begin{equation}\nonumber
\widehat{\bM}_{\text{LRE}} = \left(\frac{1}{K}\sum_{k=1}^{K}\widehat{\tau}_k\right) \widehat{\bSigma} + \bI,
\end{equation}
where $\widehat{\tau}_k$ is the estimated texture of the $k$-th clutter datum and $\widehat{\bSigma}$ is the covariance estimate of the speckle obtained through the iterative algorithm proposed in \cite{SunPalomar}.
}
\end{itemize}
Inspection of Fig. \ref{fig_compound_spatial} reveals that FNE and SNE are basically equivalent and outperform in terms of average SINR all the counterparts, included those specific for compound Gaussian disturbance\footnote{As to the LRE, the clutter covariance matrix rank is evaluated as the number of the eigenvalues greater than $\tr(\bM_s)/10^5\geq 10^{-4}$. Hence, 
the estimators exploiting the true rank, i.e., 6
(LRE-6) and  rank 7 (LRE-7) are displayed.}. Precisely, in the matched condition, as already observed in Fig. \ref{fig_gauss_spatial}, FNE, SNE, FML, and CML are almost coincident. Besides, compared to NSCM, FPE, LRE-6, LRE-7, and SCM, they  provide SINR gains up to: $11$ dB, $11$ dB, $1.89$ dB,  $1.90$ dB, and $6.93$ dB for $K=4$; $6.27$ dB, $6.27$ dB, $4.74$ dB, $7.50$ dB, and $5.38$ dB for $K=8$;  $4.77$ dB, $2.01$ dB, $6.65$ dB, $9.20$ dB, $1.63$ dB for $K=16$. Without surprise, LRE-6 outperforms LRE-7 reflecting the presence of a mismatch loss. Finally, in the mismatched scenario, FNE and SNE also grant better performance than FML and CML with gains of $1.35$ dB and $0.94$ dB for $K=4$, $2.35$ dB and $0.84$ dB, for $K=8$,  $1.51$ dB and $0.42$ dB, for $K=16$, clearly highlighting the effectiveness of the new devised strategies. As to the comparisons with the other estimators, considerations similar to those for the matched scenario holds true.

\subsection{Doppler Processing}
\begin{figure*}[ht!]
\begin{center}
\subfigure[$\sigma_a^2=0$ dB, $K=8$]{\includegraphics[width=0.49\textwidth]{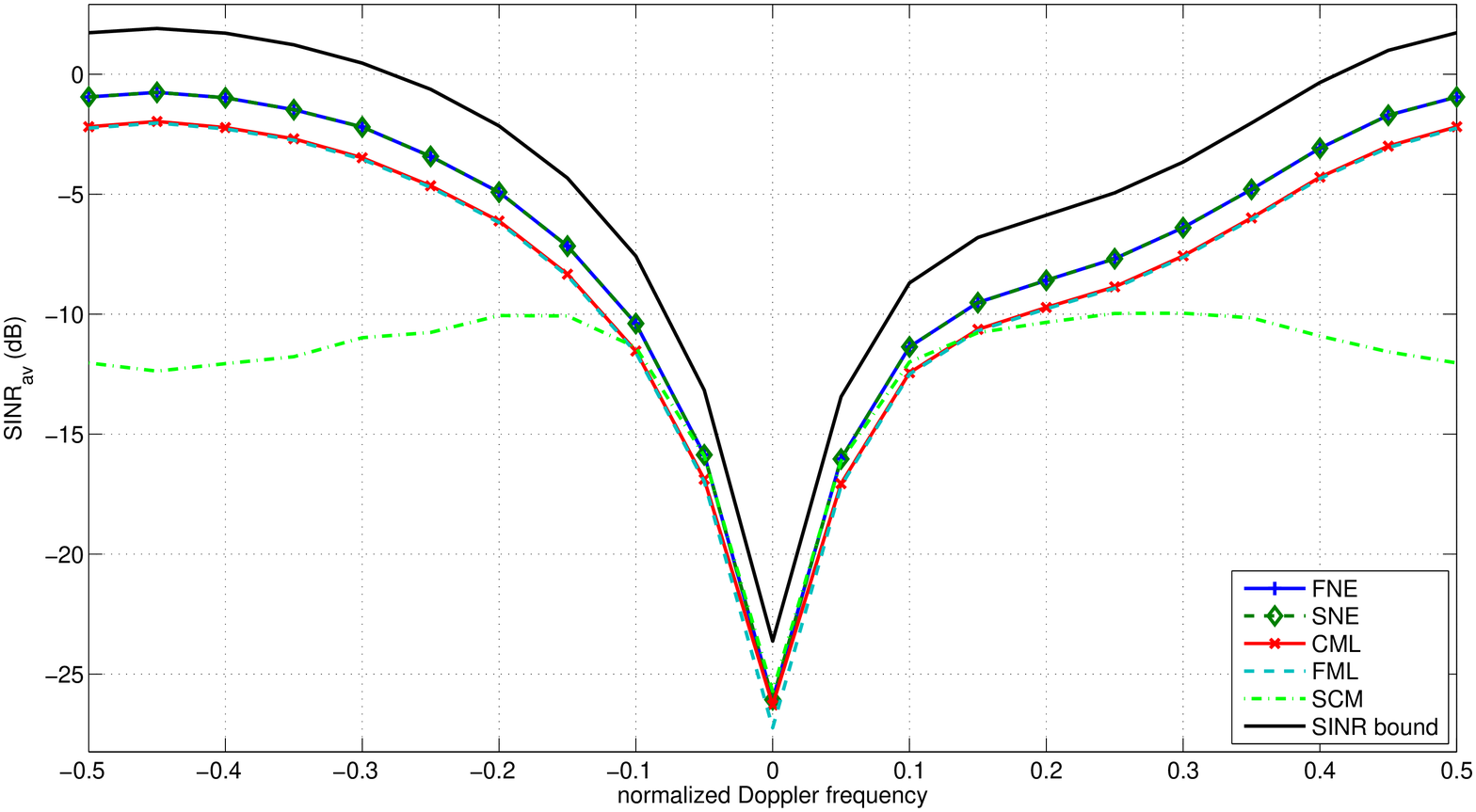}}
\subfigure[$\sigma_a^2=10$ dB, $K=8$]{\includegraphics[width=0.49\textwidth]{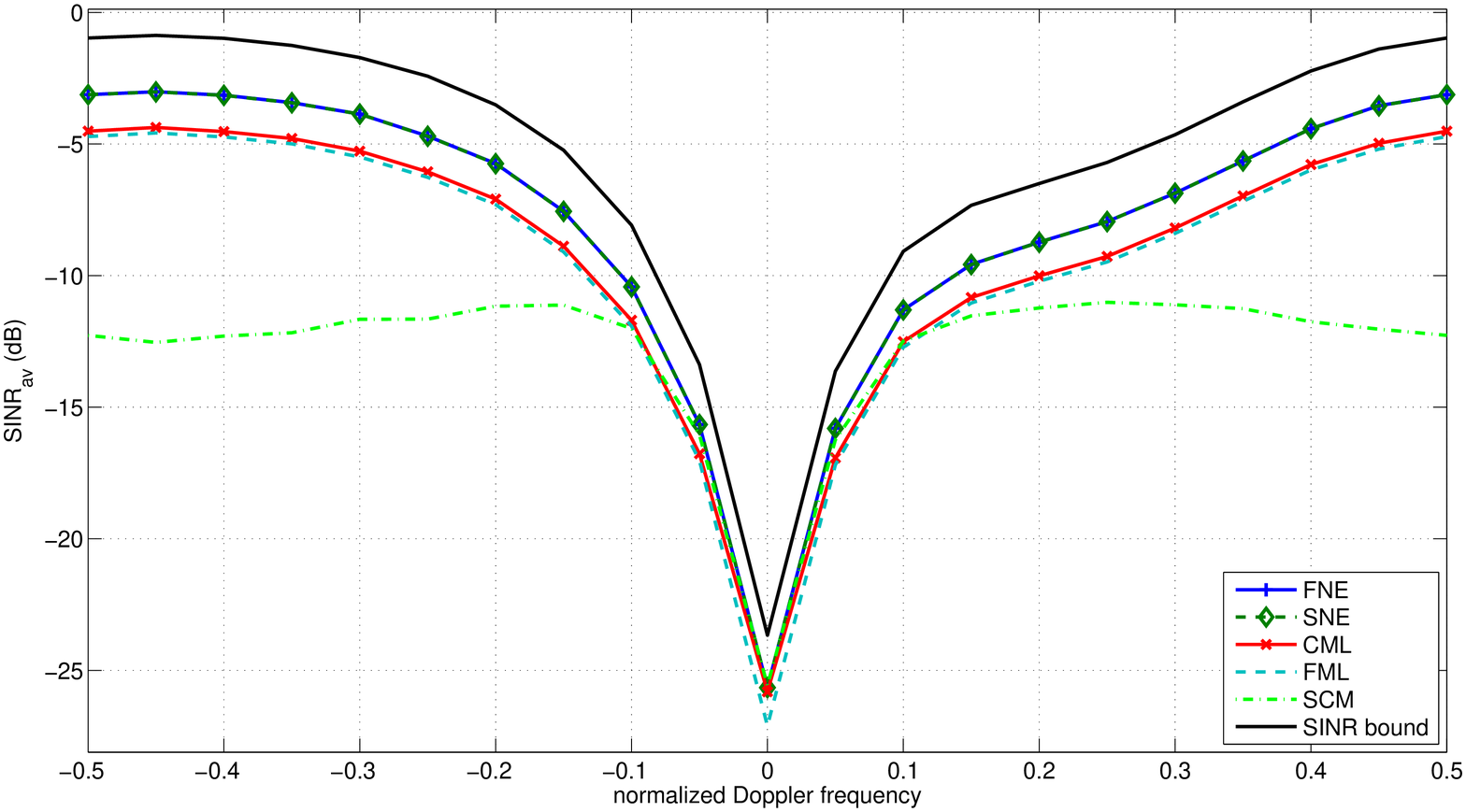}}
\subfigure[$\sigma_a^2=0$ dB, $K=16$]{\includegraphics[width=0.49\textwidth]{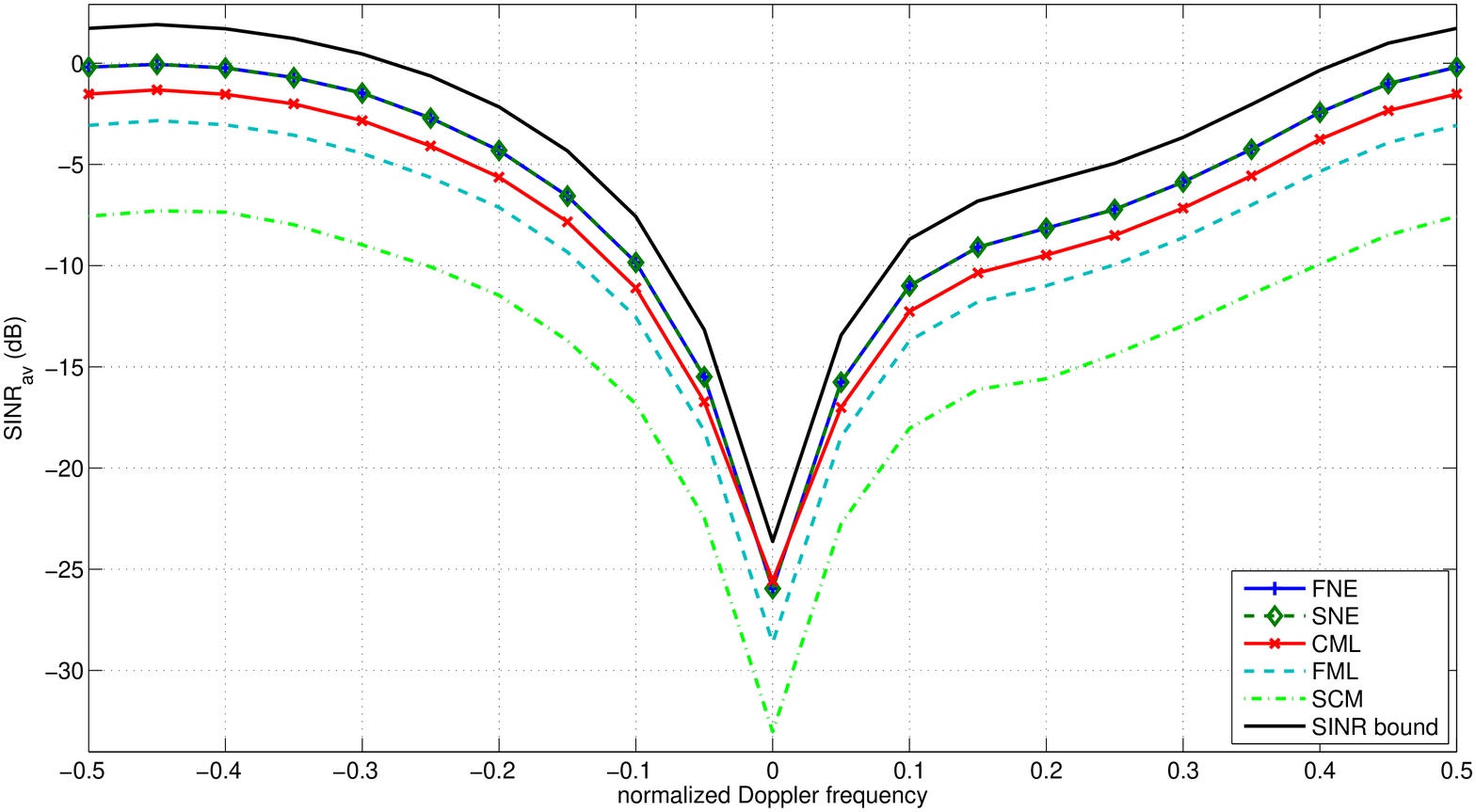}}
\subfigure[$\sigma_a^2=10$ dB, $K=16$]{\includegraphics[width=0.49\textwidth]{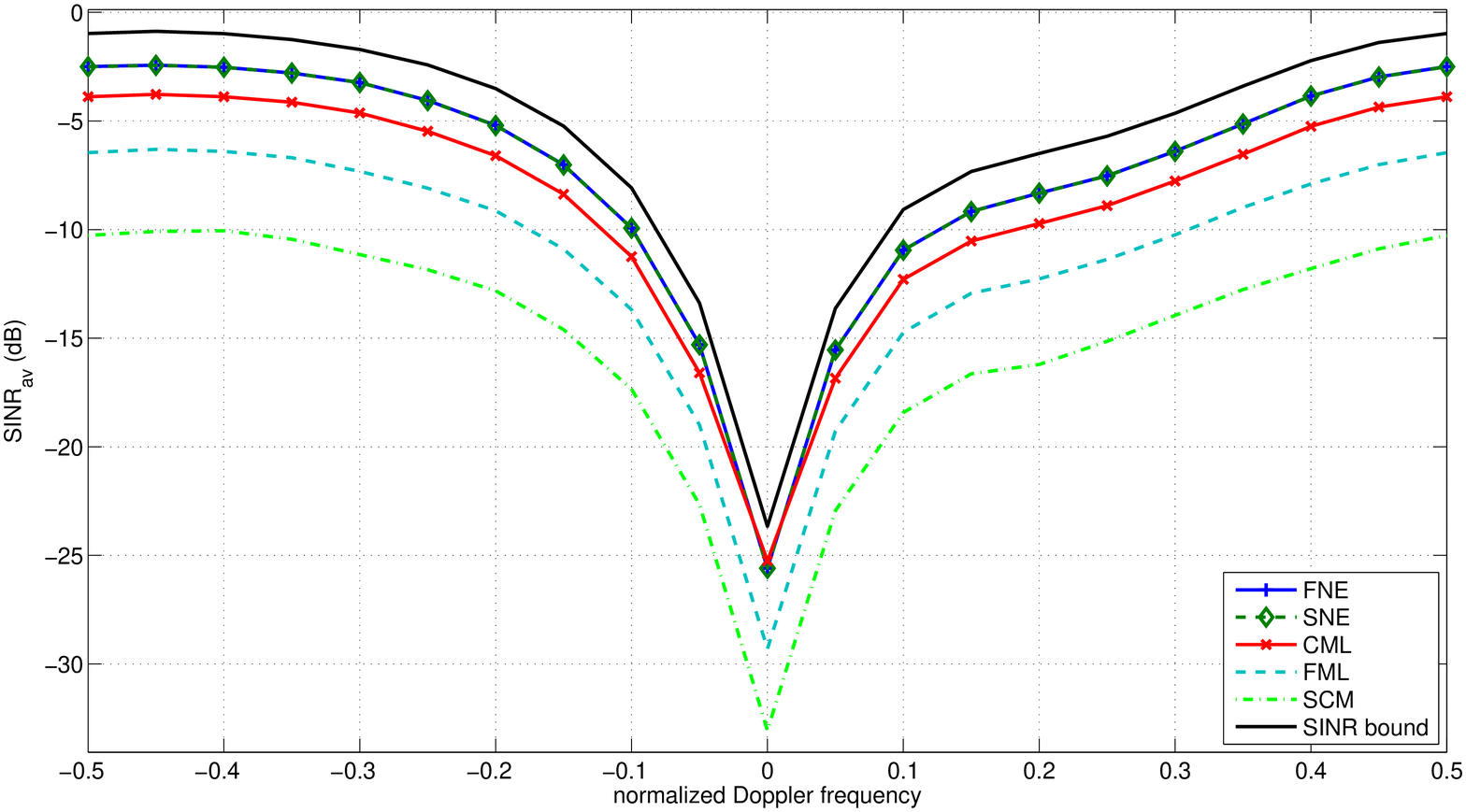}}
\subfigure[$\sigma_a^2=0$ dB, $K=32$]{\includegraphics[width=0.49\textwidth]{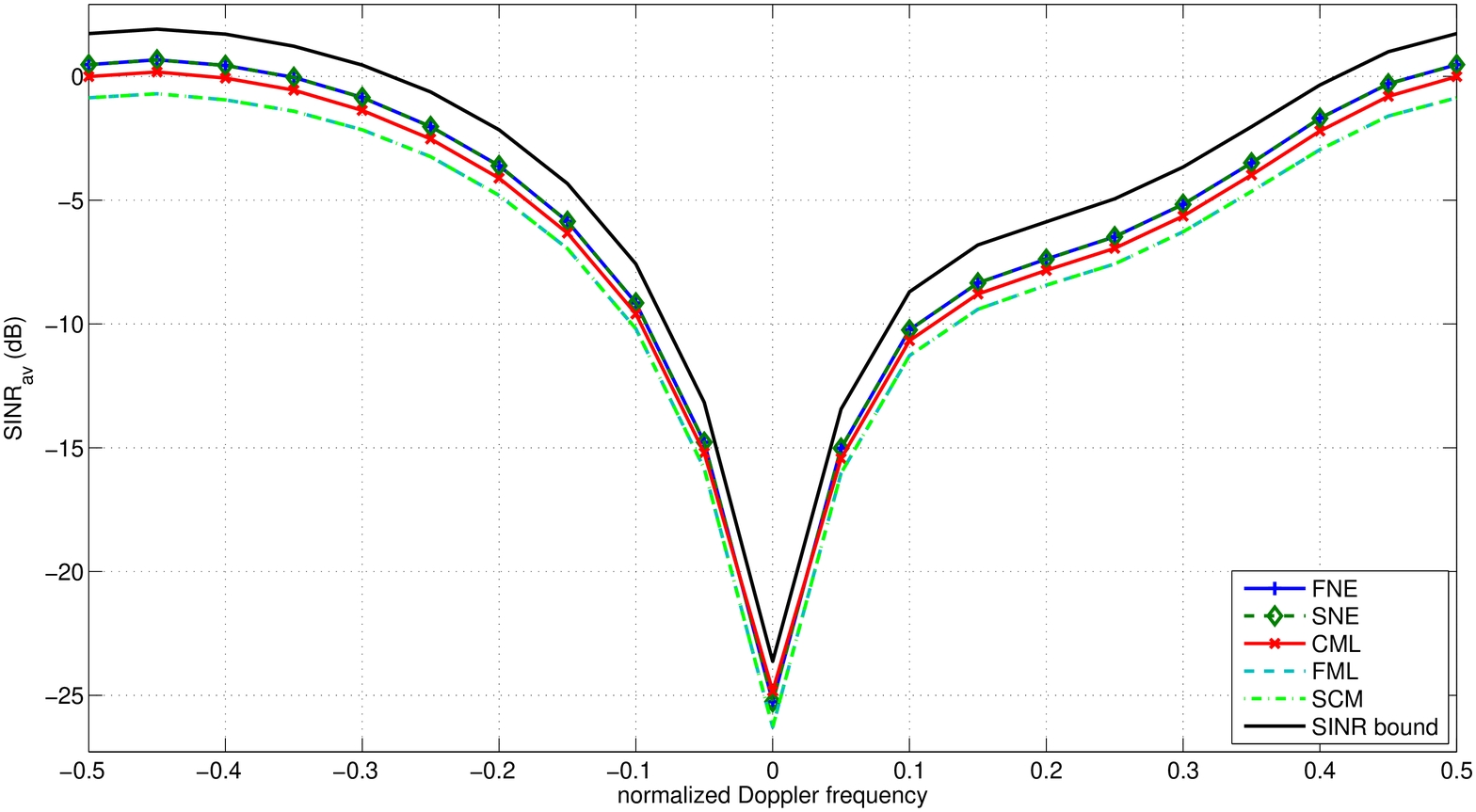}}
\subfigure[$\sigma_a^2=10$ dB, $K=32$]{\includegraphics[width=0.49\textwidth]{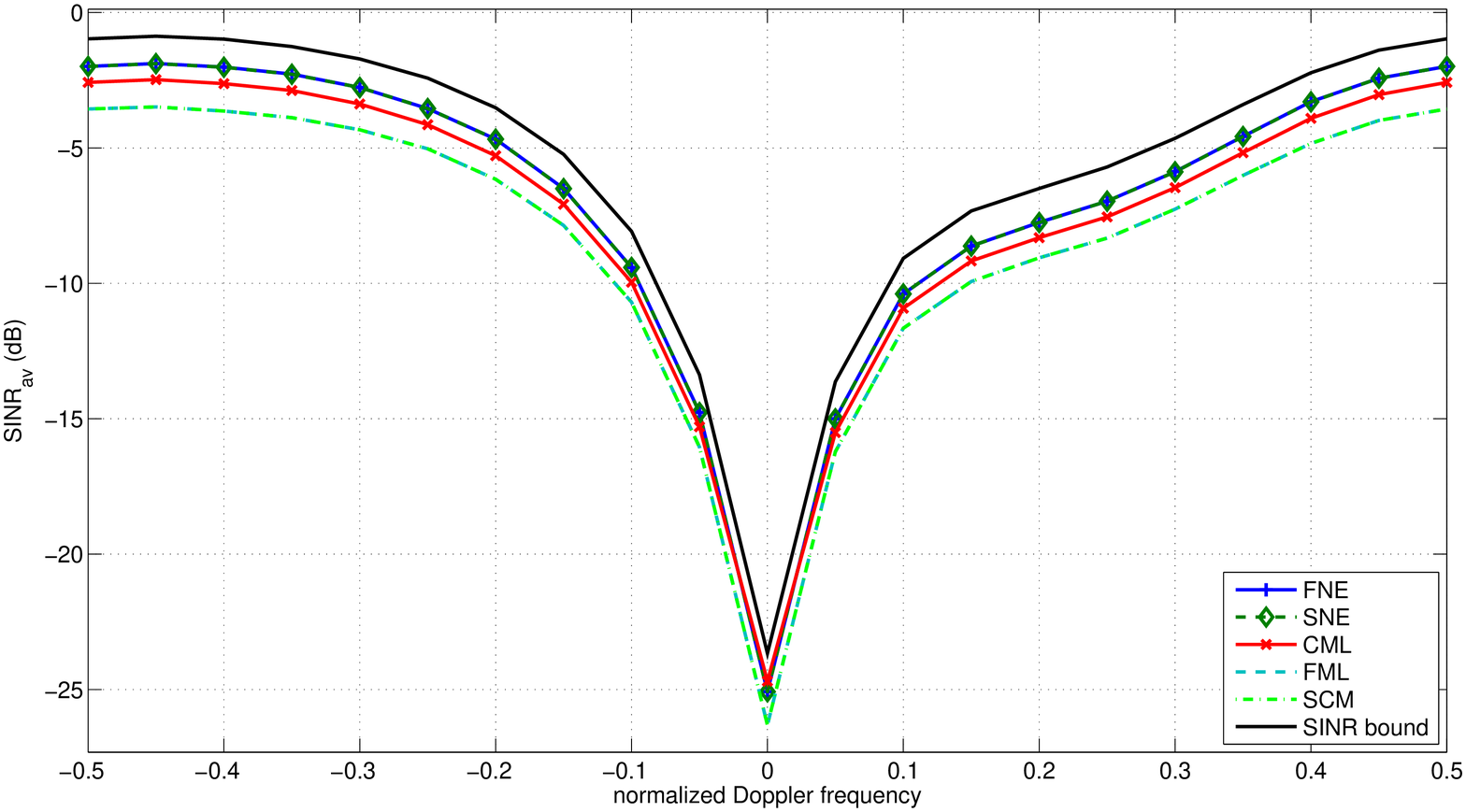}}
\end{center}
\caption{Doppler processing in the presence of Gaussian data for different numbers of training data. $\mbox{SINR}_{av}$ versus normalized Doppler frequency (blue curve $+$-marked FNE, green dashed curve SNE, red $\times$-marked curve CML, cyan dashed curve FML, green dot-dashed curve SCM, and black curve ideal SINR bound). The analyzed environment comprises a bimodal clutter composed of sea and ground clutter with $\text{CNR}_S = 10$ dB, $\text{CNR}_G = 25$, $\rho_S = 0.8$, $\rho_G = 0.95$, and $f_S = 0.2$. Subplots (a), (c), and (e) refer to the case $\sigma_a^2=0$ dB, whereas subplots (b), (d), and (f) to  $\sigma_a^2=10$ dB. \label{fig_gauss_Doppler}}
\end{figure*}

A radar system transmitting a coherent burst of $N = 16$ pulses is considered. In this case  $x$ refers to the normalized Doppler frequency of the target, i.e. $\nu \in [-1/2,1/2[$ and  the  steering vector in \eqref{sinr} reduces to 
\begin{eqnarray}\nonumber
\bs(\nu)=\left[1, \exp\left(j 2 \pi \nu \right), \ldots, \exp\left(j 2 \pi \nu (N-1)\right)\right]^T\in \mathbb{C}^N.
\end{eqnarray}
 As to the interference environment, it is assumed that the radar operates in the presence of both ground and sea clutter in addition to white noise \cite{farina1996improvement}. Therefore the overall disturbance covariance matrix is 
\begin{equation}
\bM = \bM_t + \sigma_a^2 \bI,
\end{equation}
where, $\forall\left(n,m\right)\in\left\{1,\ldots,N\right\}^2$, 
\begin{equation}\label{cov_dopp}
\begin{split}
\bM_t(n, m) = &\text{CNR}_S \rho_S^{(n-m)^2} e^{j2\pi(n-m)f_S} + \text{CNR}_G \rho_G^{|n-m|},\\
\end{split}
\end{equation}
with 
\begin{itemize}
\item $\text{CNR}_S$ and $\text{CNR}_G$  the power of the sea and ground clutter, respectively; 
\item $\rho_S$ and $\rho_G$ the one-lag correlation coefficients of the sea and the ground clutter, respectively;
\item $f_S$  the normalized Doppler frequency of the sea clutter.
\end{itemize}
In Fig. \ref{fig_gauss_Doppler}, the average SINR versus $\nu$ is reported for FNE, SNE, FML, CML, and SCM estimators. The training data are drawn from a complex circular Gaussian distribution and the parameters in (\ref{cov_dopp}) are $\text{CNR}_S = 10$ dB, $\text{CNR}_G = 25$ dB, $\rho_S = 0.8$, $\rho_G = 0.95$, and $f_S = 0.2$. Figs. \ref{fig_gauss_Doppler}(a), \ref{fig_gauss_Doppler}(c), and \ref{fig_gauss_Doppler}(e) refer to $K=8$, $K=16$, and $K=32$, respectively and assume $\sigma_a^2=0$ dB. The mismatched analysis, i.e., $\sigma_a^2=10$ dB, is instead reported in Figs. \ref{fig_gauss_Doppler}(b), \ref{fig_gauss_Doppler}(d), and \ref{fig_gauss_Doppler}(f).

The plots clearly illustrate the effectiveness of the new devised estimators. Indeed, both in the matched and in mismatched scenario, FNE and SNE achieve higher SINR values than the counterparts at each target Doppler frequency. Interestingly, unlike the spatial-processing, in this case FNE and SNE also outperform CML and FML  in the matched conditions. Precisely, the SINR gains with respect to the CML, that is the major competitor, are, for $K=8,16,32$, respectively: $1.29$ dB, $1.38$ dB, and $0.52$ dB in the matched case and $1.43$ dB, $1.42$ dB, and $0.61$ dB in mismatched situation.
As expected, the gains are lower and lower as $K$ increases, due to the consistency of all the involved estimators.
 
\section{Conclusions}\label{conclusioni}

The disturbance covariance matrix estimation problem for radar signal processing applications has been addressed according to a geometric approach. Specifically, a new family of distribution-free covariance estimators has been introduced performing the sample covariance matrix  projection, according to a specific unitary invariant norm, into a structured covariance set of practical relevance.  

To tackle the resulting constrained optimization problem an efficient solution technique has been designed which represents the main technical contribution of this paper. In particular, it has been proved that each estimator exhibits a shrinkage-type form with the eigenvalues estimate obtained via the solution of a one-dimensional convex problem tuned to the considered unitary norm. Furthermore,  almost closed form covariance estimates have been provided assuming either Frobenius or spectral norms at the design stage. Remarkably, the proposed estimators possess the consistency property as long as the training vectors are statistically independent. 

Some interesting case studies have been considered to illustrate the effectiveness of the new proposed framework. The results have shown that the new estimators may provide better SINR values than some structured estimators available in the open literature. Precisely, the lower the sample support the higher the gain. 
 Additionally, accounting for
the computational efforts as well as the SINR performance, the
best estimator appears that based on spectral norm.

As possible future research tracks, it might be worth analyzing the performance of the new family of estimators on real radar data as well as to account for other unitary invariant constraint at the design stage such as an upper bound on the clutter rank.
\appendix


\subsection{Proof of Lemma \ref{teorema_2_lambda}}\label{appendice_teorema_2_lambda}
\begin{proof}
Let ${\bX}_{1}^{\star}={\bU}_1{\bLambda}_1^{\star}{\bU}^{\dag}_1$ be the spectral decomposition of an optimal solution to ${\cal{P}}_1$, where ${\bLambda}_1^{\star}=\diag(\lambda_1^{\star},\lambda_2^{\star}, \cdots,\lambda_N^{\star})$, with $\lambda_1^{\star}\geq\lambda_2^{\star}\geq \cdots\geq\lambda_N^{\star}$. Based on  \cite[Theorem 7.4.51]{Horn}
\begin{equation}\label{eq_lower_bound}
\|{\bX}_{1}^{\star} - \bS\|\geq \|{\bLambda}_1^{\star} - {\bLambda}_S\|.
\end{equation}
Besides,
\begin{equation}
\|{\bLambda}_1^{\star} - {\bLambda}_S\|=\|{\bU}_S({\bLambda}_1^{\star} - {\bLambda}_S){\bU}_S^{\dag}\|=\|\bX^{\star} - \bS\|,
\end{equation}
due to the unitary invariance of the norm. Hence, $\bX^{\star}={\bU}_S{\bLambda}_1^{\star}{\bU}_S^{\dag}$ is an optimal solution to ${\cal{P}}_1$ since it is a feasible point achieving the minimum value. Finally, ${\cal{P}}_2$ is obtained replacing $\bX$ with ${\bU}_S{\bLambda}{\bU}_S^{\dag}$ in ${\cal{P}}_1$.  
\end{proof}

\subsection{Proof of Lemma \ref{lemma_soluzione_lambda(u_segnato)}}\label{appendice_lemma_soluzione_lambda(u_segnato)}
\begin{proof}
Let $g(\cdot)$ be the gauge function associated with the considered unitary invariant norm $\|\cdot\|$ (see \cite[Theorem 7.4.24]{Horn}). According to  \cite[Property 7.4.21]{Horn},
$$\|\bLambda-\bLambda_S\|=g(|\lambda_1-d_1|,|\lambda_2-d_2|,\ldots, |\lambda_N-d_N|),$$
implying that ${\cal{P}}_2^{\prime}({u})$ can be cast as 

\begin{equation}\nonumber
\left\{
\begin{array}[c]{ll}
\underset{\blambda}{\min} & g(|\lambda_1-d_1|,|\lambda_2-d_2|,\ldots, |\lambda_N-d_N|)\\
\mbox{s.t.} & \lambda_i\geq1, \quad i=1,\ldots,N\\
 & {u}\leq{\lambda_i}\leq{\kappa_M {u}},\quad i=1,\ldots,N\\
 & {u}\geq{\frac{1}{\kappa_M }}
\end{array}
\right..
\end{equation}

To proceed further, let us observe that $g(\cdot)$ is a monotone norm on $\mathbb{C}^N$ \cite[Theorem 5.5.10]{Horn}. As a consequence, given two vectors $\bx\in\mathbb{C}^N$ and $\by\in\mathbb{C}^N$ such that $|x_i|\leq |y_i|, i=1,\ldots,N$,  ${g(\bx)}\leq{g(\by)}$. 
 Additionally,  for any fixed $u=\bar{u}$ the constraints on the variables $\lambda_i$, $i=1,\ldots,N$ are not coupled. Thus, an optimal solution to 
\begin{equation}\nonumber
\left\{
\begin{array}[c]{ll}
\underset{\blambda}{\min} & g(|\lambda_1-d_1|,|\lambda_2-d_2|,\ldots, |\lambda_N-d_N|)\\
\mbox{s.t.} & \lambda_i\geq1, \quad i=1,\ldots,N\\
 & \bar{u}\leq{\lambda_i}\leq{\kappa_M \bar{u}},\quad i=1,\ldots,N\\
\end{array}
\right.
\end{equation}
can be found solving the following $N$ scalar optimization problems
\begin{equation}\label{appendice_problema_Pprimo3(u)}
{\cal{P}}_3^i(\bar{u})\left\{
\begin{array}[c]{ll}
\underset{\lambda_i}{\min} & |\lambda_i-d_i|\\
\mbox{s.t.} & \lambda_i\geq1\\
\ & \overline{u}\leq{\lambda_i}\leq{\kappa_M\bar{u}}\\
\end{array},
\right.
\quad i=1,\ldots,N.
\end{equation}

The closed form solution to  ${\cal{P}}_3^i(\bar{u})$ can be obtained analyzing 
\begin{equation}
\tilde{{\cal{P}}}_3\left\{
\begin{array}[c]{ll}
\underset{x}{\min} & |x-y|\\
\mbox{s.t.} & a\leq{x}\leq{b}\\
\end{array}
\right.,
\end{equation}
where the variables $x, y, a, b$ are given by
\begin{equation}\label{appendice_sostituzioni_x,y,a,b}
x=\lambda_i,\quad y=d_i,\quad a=\max(1,\bar{u}),\quad b=\kappa_M \bar{u}.
\end{equation}
Since $|x-y|$ is a monotonically decreasing function if $x\leq y$ and a monotonically increasing function if 
$x\geq y$, it follows that the optimal solution is $x_{min}=y$ if $a \leq y \leq b$. If $y \leq a\leq b$ $x_{min}=a$. Thus,  $x_{min}=\max(y,a)$ as long as $y\leq b$. Moreover, $x_{min}=b$ if $y\geq b$  implying that the optimal solution to $\tilde{{\cal{P}}}_3$ can be written in closed form as
\begin{equation}\label{ottimizzata}
x_{min}=\min(b,\max(y,a)).
\end{equation}
Replacing \eqref{appendice_sostituzioni_x,y,a,b} in (\ref{ottimizzata}), the optimal solution to ${\cal{P}}_3^i(\bar{u})$ is
$$
\lambda^{\star}_i(\bar{u})={\min}(\kappa_M\bar{u}, \max(d_i,\max(1,\bar{u}))).
$$
\end{proof}

\subsection{Proof of Theorem \ref{espressione_finale}}
\label{espressione_finale_prrof}
\begin{proof}
Let $\bLambda^{\star}(\bar{u})$ be an optimal solution to problem ${\cal{P}}_2^{\prime}(\bar{u})$ in \eqref{equation_variabile_u} as provided by Lemma \ref{lemma_soluzione_lambda(u_segnato)}. Hence,  a minimizer of ${\cal{P}}_2$ is $\bLambda^\star=\bLambda^{\star}(u_1^{\star})$ with $u_1^\star$  an optimal solution to the one-dimensional problem
\begin{equation}
{\cal{P}}_3^{\prime}\left\{
\begin{array}[c]{ll}
\underset{u}{\min} & \|\bLambda^{\star}(u) - \bLambda_S\|\\
\mbox{s.t.} & u\geq \frac{1}{\kappa_M}
\end{array}
\right..
\end{equation}
Based on \cite[Property 7.4.21]{Horn} ${\cal{P}}_3^{\prime}$  can be reformulated as
\begin{equation}
{\cal{P}}_3\left\{
\begin{array}[c]{ll}
\underset{u}{\min} &  g\left(|\lambda_1^{\star}(u)-d_1|,\ldots,|\lambda_N^{\star}(u)-d_N|\right)\\
\mbox{s.t.} & u\geq \frac{1}{\kappa_M}
\end{array}
\right.,
\end{equation}
where  $g(\cdot)$ is the gauge function associated with the considered unitary invariant norm and $\lambda_i^{\star}(u)$, $i=1,\ldots,N$, are the entries of the vectorial function $\blambda^{\star}(u)$ given in \eqref{espressione_di_lambda(u)}.  Now, notice that, $g\left(|\lambda_1^{\star}(u)-d_1|,\ldots,|\lambda_N^{\star}(u)-d_N|\right)$ is a continuous function since obtained as the composition of continuous functions. Besides, due to \cite[p. 88]{ConvexOptimization_Boyd}, the objective function in ${\cal{P}}_3$ is convex. Hence, an optimal solution to ${\cal{P}}_1$ just requires the solution of the convex optimization problem ${\cal{P}}_3$. Additionally,  the set of the optimal solutions to ${\cal{P}}_3$ defines a closed convex interval bounded below ensuring the existence of the lowest optimal solution $u^{\star}$.
\end{proof}

\subsection{Proof of Theorem \ref{teorema_soluzione_d1_magg_Kmax}}
\label{appendice_teorema_soluzione_d1_magg_Kmax}
\begin{proof}
The proof is organized in four different parts accordingly to the provided claims.
\begin{enumerate}
\item As proved in the supplementary material, $G_1(u)$ is a convex and differentiable function within the interval $u\in [1,d_1]$. Recalling that a differentiable function is convex if and only if its derivative is an increasing function, it can be claimed that $\frac{d G_1(u)}{d u}$ is increasing within  $u\in [1,d_1]$. Since $\frac{d G_1(u)}{d u}\Big|_{u=1} \geq 0$, then
$$
\frac{d G_1(u)}{d u}\geq 0,\quad \quad u\in [1,d_1],
$$
namely $G_1(u)$ is an increasing function in $u\in [1,d_1]$. Hence, the minimum to ${\cal{P}}_4$ is attained in $u^{\star} = 1$.

\item Assume that
\begin{equation}\label{appendice_soluz_d1>Kmax_caso2}
\frac{d G_1(u)}{d u}\Big|_{u=1}< 0 \quad \mbox{and} \quad \frac{d G_1(u)}{d u}\Big|_{u=d_1}\leq 0.
\end{equation}
Since, $G_1(u)$ is a convex and differentiable function within the interval $u\in [1,d_1]$,
$$
\frac{d G_1(u)}{d u}\leq 0, \quad\quad u\in [1,d_1],
$$
therefore $G_1(u)$ is decreasing in $[1,d_1]$ and $u_1^{\star} = d_1$ is the optimal solution as long as\footnote{Without loss of generality, it is assumed that $v_i\neq v_j$ and $v_i\neq \kappa_M  v_j$ for all $i\neq j$ with $1\leq i, j\leq \bar{N}+1$.}
 $d_1> d_N$. 
Indeed, if there exists $u^{\star} < d_1$,  $G_1(u)$ would be constant over $[u^{\star}, u_1^{\star}]$ but this  is not possible since  $d_1> d_N$.
\item If $\frac{d_1}{\kappa_M}\leq d_N$, $G_1(u)$ is strictly decreasing up to $\frac{d_1}{\kappa_M}$ and increasing after this point implying that  $u^{\star}=\frac{d_1}{\kappa_M}$.
\item Finally, if $\frac{d_1}{\kappa_M} > d_N$, the optimal solution is unique. Indeed, if $G_1(u)$ is constant over an interval its second order derivative is null in this set. Now, observe that $G_1(u)$ is composed of convex functions whose second order derivative exists except for a finite number of points. Additionally, in any regular interval there is at least one of such functions that is strictly convex due to the assumption $\frac{d_1}{\kappa_M} > d_N$, and hence the initial claim is contradicted. Now, owing to
$$
\frac{d G_1(u)}{d u}\Big|_{u=1}< 0 \quad \mbox{and} \quad \frac{d G_1(u)}{d u}\Big|_{u=d_1} > 0,
$$
the optimal point belongs to $]1,d_1[$. Since $G_1(u)$ is a convex and differentiable function, $u^{\star}$ is the optimal point if and only if
\begin{equation}\label{appendice_condiz_necess_suffic_per_u_star}
\frac{d G_1(u)}{d u}\Big|_{u=u^{\star}} = 0.
\end{equation}

Let us now fully characterize condition  \eqref{appendice_condiz_necess_suffic_per_u_star}. To this end, let $\alpha \in \{1,2,\ldots,\bar{N},\bar{N}+1\}$ be the smallest index such that $v_\alpha<u^{\star}$ ($\alpha\geq 2$ since $u^{\star} < d_1 = v_1$; moreover $\alpha\leq \bar{N}+1$ because $u^{\star} > 1 = v_{\bar{N}+1}$). Besides, let $\beta \in \{1,2,\ldots,\bar{N},\bar{N}+1\}$ be the largest index such that $\frac{v_\beta}{\kappa_M }>u^{\star}$ ($\beta\geq 1$ because\footnote{$\frac{d G_1(u)}{d u}\Big|_{u=\frac{d_1}{\kappa_M }}>0$ since each term composing $G_1(u)$ is an increasing function over $u \geq \frac{d_1}{\kappa_M }$ and at least the term associated with $d_N$ has a strictly positive derivative.} $u^{\star}< \frac{d_1}{\kappa_M }$ and $\beta\leq \bar{N}$ since $u^{\star}> 1 > \frac{1}{\kappa_M } = \frac{v_{\bar{N}+1}}{\kappa_M }$). Notice that, $\alpha > \beta$, otherwise $v_\alpha\geq v_\beta > \frac{v_\beta}{\kappa_M } > u^{\star}$, which contradicts $u^{\star} > v_\alpha$. Therefore, it will exist a neighborhood  $B_{u{^\star}}$ of $u^{\star}$, contained within the interval $[v_\alpha, \frac{v_\beta}{\kappa_M }]$, such that $G_1(u)$ can be expressed as
\begin{equation}\label{appendice_espressione_G(u)_nell'intervallo_ottimo}
G_1(u) = \displaystyle{\sum_{i=1}^{\beta}} \left(d_i - \kappa_M u \right)^2 + \displaystyle{\sum_{i=\alpha}^{N}} \left(u - d_i \right)^2, \; \forall\, u \in B_{u^{\star}}.
\end{equation}
Computing the derivative of \eqref{appendice_espressione_G(u)_nell'intervallo_ottimo} with respect to $u$, and imposing the optimality condition \eqref{appendice_condiz_necess_suffic_per_u_star}, the minimizer $u^{\star}$ is given by
\begin{equation}
u^{\star} = \frac{\displaystyle{\sum_{i=1}^{\beta}} \kappa_M d_i + \displaystyle{\sum_{i=\alpha}^{N}} d_i}{N - \alpha + 1 + \beta \kappa_M ^2}.
\end{equation}
On the other hand, let $\bar{u}$ a value such that
\begin{equation}\label{appendice_u_segnato_con_alfa_e_beta}
\bar{u} = \frac{\displaystyle{\sum_{i=1}^{\bar{\beta}}} \kappa_M d_i + \displaystyle{\sum_{i=\bar{\alpha}}^{N}} d_i}{N - \bar{\alpha} + 1 + \bar{\beta} \kappa_M ^2},
\end{equation}
with $\bar{\alpha} \in \{1,2,\ldots,\bar{N},\bar{N}+1\}$ the smallest indexes such that $v_{\bar{\alpha}} < \bar{u}$ and $\bar{\beta} \in \{1,2,\ldots,\bar{N},\bar{N}+1\}$ the largest indexes such that $\frac{v_{\bar{\beta}}}{\kappa_M } > \bar{u}$. This means that it exists a neighborhood $B_{\bar{u}}$ of $\bar{u}$ contained within the interval $\left[v_{\bar{\alpha}}, \frac{v_{\bar{\beta}}}{\kappa_M }\right]$, such that $\forall u \in B_{\bar{u}}$, $G_1(u)$ is given by
\begin{equation}\label{appendice_espress_G(u_segnato)_nell'intervallo_ottimo}
G_1(u) = \displaystyle{\sum_{i=1}^{\bar{\beta}}} \left(d_i - \kappa_M  u \right)^2 + \displaystyle{\sum_{i=\bar{\alpha}}^{N}}\left(u - d_i \right)^2.
\end{equation}
Now, computing the derivative of \eqref{appendice_espress_G(u_segnato)_nell'intervallo_ottimo}, condition \eqref{appendice_u_segnato_con_alfa_e_beta} implies that
$ \frac{d G_1(u)}{d u}\Big|_{u = \bar{u}}= 0$, i.e., $\bar{u}=u^{\star}$.
\end{enumerate}
\end{proof}

\subsection{Proof of Theorem \ref{sp_funzione_G(u)}}
\label{appendice_sp_funzione_G(u)}
\begin{proof}
The proof is organized in five different parts accordingly to the provided claims.
\begin{enumerate}
\item If $d_i \leq 1$, $i=1,\ldots,N$, the functions $h_i(u)$, $i=1,\ldots,N$, in (\ref{appendice_h(u)caso_d_i<1})
 are monotonically increasing over $u \geq \frac{1}{\kappa_M}$.
 Hence, $G_2(u)$ is monotonically increasing implying that $u^{\star} = \frac{1}{\kappa_M}$.
\item Let $\mathcal{I}_q=\{i\,:\,\, d_i \leq 1\}$ and $\mathcal{I}_p=\{i\,:\,\, d_i>1\}$; by assumption, $1 < d_1 \leq \kappa_M $ and $\mathcal{I}_q\neq \emptyset$. Now, observe that 
\begin{equation}\nonumber
G_2(u)=\underset{i=1,\ldots,N}{\max}\left\{h_i(u)\right\}={\max}\left\{h^q(u),h^p(u)\right\},
\end{equation}
 where 
\begin{equation}\nonumber
h^q(u)=\underset{i\in \mathcal{I}_q}{\max}\left\{h_i(u)\right\},\quad h^p(u)=\underset{i\in \mathcal{I}_p}{\max}\left\{h_i(u)\right\}.
\end{equation}
To proceed further, notice that
\begin{equation}\nonumber
h^p(u) = \left\{
\begin{array}[c]{ll}
h_1(u) & \mbox{if}\quad \frac{1}{\kappa_M }\leq u\leq 1\\
h_{p_{max}}(u) & \mbox{if}\quad 1\leq u\leq d_1\\
\end{array}
\right.
\end{equation}
where $p_{max}=\max\{j\in \mathcal{I}_p\}$. In fact,  
\begin{equation}\nonumber
\max\left\{d_1-\kappa_M u,0\right\}\geq \max\left\{d_j-\kappa_M u,0\right\},\,\,j\in \mathcal{I}_p,
\end{equation}
implying that 
\begin{equation}\nonumber
h^p(u)=h_1(u)\,\, \mbox{over}\,\, \left[\frac{1}{\kappa_M},1\right]\subseteq \left[\frac{1}{\kappa_M},d_{p_{max}}\right];
\end{equation}
moreover, $\forall j\in \mathcal{I}_p$, $\frac{d_j}{\kappa_M }<1$ and
\begin{equation}\nonumber
\max\left\{u-d_{p_{max}},0\right\}\geq \max\left\{u-d_j,0\right\},
\end{equation}
thus $h^p(u)=h_{p_{max}}(u)$ over $u \geq 1$. Finally, since $h^q(u)=h_N(u)$ if $u \geq \frac{1}{\kappa_M }$,
\begin{equation}\nonumber
G_2(u) = \left\{
\begin{array}[c]{ll}
{\max}\left\{h_N(u),h_1(u)\right\} & \mbox{if}\quad \frac{1}{\kappa_M }\leq u\leq 1\\
{\max}\left\{h_N(u),h_{p_{max}}(u)\right\} & \mbox{if}\quad 1\leq u \leq d_1\\
\end{array}
\right..
\end{equation}
To proceed further, observe that 
\begin{equation}\nonumber
{\max}\left\{h_N(u),h_{p_{max}}(u)\right\}=h_N(u) \quad u\geq 1,
\end{equation}
since $u-d_N \geq \max\left\{u-d_{p_{max}},0\right\}$, $u\geq 1$. Now, let $\eta=\frac{d_1+d_N-1}{\kappa_M }$ be the point such that $1-d_N = d_1-\eta \kappa_M $, where $\eta\leq\frac{d_1}{\kappa_M}\leq 1$ since  $d_N\leq 1$.  If $\eta\leq \frac{1}{\kappa_M }$, $G_2(u)=h_N(u)$, $u\geq \frac{1}{\kappa_M }$, implying that $u^\star=\frac{1}{\kappa_M }$. Otherwise,
\begin{equation}\nonumber
G_2(u) = \left\{
\begin{array}[c]{ll}
h_1(u) & \mbox{if}\quad \frac{1}{\kappa_M }\leq u \leq \eta\\
h_N(u) & \mbox{if}\quad  u \geq \eta\\
\end{array}
\right.
\end{equation}  
and $u^\star=\eta$ since $G_2(u)$ is a strictly decreasing function up to $\eta$ and it monotonically increases if $u\geq \eta$.  Thus, $u^\star=\max\left\{\eta,\frac{1}{\kappa_M}\right\}$.
\item Assume $1 < d_1\leq \kappa_M $ and $d_N>1$. Since, $d_i>1$, $i=1,\ldots,N$, $G_2(u)=h_1(u)$ as $\frac{1}{\kappa_M} \leq u\leq d_N$; hence,  $G_2(u)$, is a strictly decreasing function over $\frac{1}{\kappa_M} \leq u \leq \frac{d_1}{\kappa_M}$ and $G_2(\frac{d_1}{\kappa_M})=0$ since $\frac{d_1}{\kappa_M}\leq 1 <d_N$. As a consequence $u^\star=\frac{d_1}{\kappa_M}$ and  $\bX^{\star} = \bS$.
\item Consider  $d_1 > \kappa_M $ and $d_N\leq 1$. In this case,
\begin{equation}\label{espressione_compattaG_2_1}
G_2(u) \!\!=\!\! \left\{
\begin{array}[c]{ll}
\!\!\!\!{\max}\left\{h_N(u),h_1(u)\right\} \!\!\!\!& \mbox{if}\, \frac{1}{\kappa_M }\leq u\leq 1\\
\!\!\!\!{\max}\left\{h_N(u),h_1(u), h_{p_{max}}(u)\right\} \!\!\!\!& \mbox{if}\, 1\leq u\leq d_1\\
\end{array}
\right.
\end{equation}
where, as in item 2), $p_{max}$ is the index of the lowest eigenvalue higher than 1. To proceed further, let $\eta=\frac{d_1+d_N-1}{\kappa_M }$; if $\eta\leq 1$, 
\begin{equation}\label{prima_rel}
h_N(u)\geq 1-d_N\geq h_1(u),\,\,\max\left\{\eta,\frac{1}{\kappa_M}\right\}\leq u\leq d_1.
\end{equation}
Furthermore
\begin{eqnarray}\label{sec_rel}
h_1(u)&\geq& h_{p_{max}}(u),\,\,1\leq u\leq d_{p_{max}},\label{sec_rel}\\
h_N(u)&\geq& h_{p_{max}}(u),\,\,d_{p_{max}}\leq u\leq d_1\label{ter_rel}.
\end{eqnarray}
Hence, based on (\ref{prima_rel}), (\ref{sec_rel}), and (\ref{ter_rel}) 
\begin{equation}\nonumber
\max\{h_N(u),h_1(u),h_{p_{max}}(u)\}=h_1(u),\,\,1\leq u\leq d_1,
\end{equation}
implying that $u^{\star}=\max\left\{\frac{1}{\kappa_M},\eta_1\right\}$. 

Now, assume $\eta > 1$ and let $\eta_1 = \frac{d_1+d_N}{1+\kappa_M }$, where $1 \leq \eta_1 \leq \frac{d_1}{\kappa_M}$. 
According to (\ref{espressione_compattaG_2_1}), $G_2(u)=h_1(u)$, if $\frac{1}{\kappa_M}\leq u\leq 1$. Additionally, 
\begin{eqnarray}\nonumber
h_1(u)&\geq & h_N(u),\,\,1\leq u\leq \eta_1,\\
h_1(u)&\leq & h_N(u),\,\,\eta_1 \leq u\leq d_1.
\end{eqnarray}
Besides, 
\begin{itemize}
\item if $\frac{d_{p_{max}}}{\kappa_M}\leq 1$
\begin{equation}\nonumber
h_{p_{max}}(u)\leq h_N(u),\,\,1\leq u\leq d_1;
\end{equation}
\item otherwise, $\frac{d_{p_{max}}}{\kappa_M}> 1$ and
\begin{eqnarray}\nonumber
h_{p_{max}}(u)&\leq & h_1(u),\,\,1\leq u\leq \frac{d_{p_{max}}}{\kappa_M},\\
h_{p_{max}}(u)&\leq & h_N(u),\,\,\frac{d_{p_{max}}}{\kappa_M}\leq u\leq d_1.
\end{eqnarray}
\end{itemize}
Summarizing,
\begin{equation}\label{espressione_compattaG_2_2}
G_2(u) = \left\{
\begin{array}[c]{ll}
h_1(u) & \mbox{if}\quad \frac{1}{\kappa_M }\leq u\leq 1\\
{\max}\left\{h_N(u),h_1(u)\right\} & \mbox{if}\quad 1\leq u\leq d_1\\
\end{array}
\right..
\end{equation}
As a consequence, $G_2(u)$  is a strictly decreasing function over $\frac{1}{\kappa_M }\leq u\leq \eta_1$ and monotonically increases over $\eta_1 \leq u\leq d_1$. Thus, $u^\star=\eta_1$.

\item Let  $d_1 > \kappa_M $ and $d_N>1$. 
\begin{itemize}
\item If $d_N\leq{\frac{d_1}{\kappa_M }}$, let $\eta_1=\frac{d_1+d_N}{1+\kappa_M }\leq{\frac{d_1}{\kappa_M }}$. Hence, $G_2(u)=h_1(u)$, if $\frac{1}{\kappa_M }\leq u \leq \eta_1$. In fact, within this interval, $\forall i \in\{1,\ldots,N\}$ 
\begin{eqnarray}
d_1-u\kappa_M \!\! &\geq & \!\!\!\!\max\left\{d_i-u\kappa_M,0 \right\},\nonumber\\
d_1-u\kappa_M \!\! &\geq & \!\!\!\!\max\left\{u - d_N,0\right\}\geq \max\left\{ u - d_i,0\right\}.\nonumber
\end{eqnarray}
Moreover, $\eta_1\geq d_N$, since $\frac{d_1+d_N}{1+\kappa_M }-d_N=\frac{d_1/\kappa_M -d_N}{\kappa_M (1+\kappa_M )}\geq 0$ and consequently  $G_2(u)=h_N(u)$ if $\eta_1 \leq u\leq d_1$. As a result, $u^{\star} = \frac{d_1+d_N}{1+\kappa_M }$;
\item otherwise, $d_N > \frac{d_1}{\kappa_M }$ and following the same line of reasoning as in item 3) it follows that $u^\star=\frac{d_1}{\kappa_M} $ as well as $\bX^{\star} = \bS$.
\end{itemize}
\end{enumerate}
\end{proof}

\bibliographystyle{ieeetr}
\bibliography{biblio}

\clearpage
\newpage

\section*{SUPPLEMENTARY MATERIAL ORGANIZATION}
The following additional appendices contain supplementary material for this paper. More specifically, the following sections provide detailed proofs of some claims within the manuscript.

\section*{APPENDIX S1. Suffiecient condition for estimator measurability}
\begin{proof}
Let us show that the estimator $\widehat{\bM}$ is a continuous (thus measurable) function of $\widehat{\bS}$ if $\|\cdot\|$ is strictly convex. To this end, let $\widehat{\bS}_i \in \mathbb{H}^N$, $i=1,2\ldots$ be a sequence of positive semi-definite matrices converging to $\widehat{\bS}^\star$ as $i \rightarrow \infty$ (e.g., $\widehat{\bS}_i\rightarrow \widehat{\bS}^\star$) and $\widehat{\bM}_i$ the resulting sequence of covariance estimates. Now, the goal is to show that $\widehat{\bM}_i\rightarrow \widehat{\bM}^\star$ with $\widehat{\bM}^\star$ the estimate associated with $\widehat{\bS}^\star$. Based on \cite[Lemma IV.1]{robust_huang}, $\displaystyle{\lim_{i \rightarrow \infty}}\displaystyle{\min_{\widehat{\bM}\in \mathcal{M}}}\|\widehat{\bS}_i - \widehat{\bM}\|=\displaystyle{\min_{\widehat{\bM}\in \mathcal{M}}}\|\widehat{\bS}^\star - \widehat{\bM}\|$, where $\mathcal{M}$ is a compact set contained in (3) that encompasses the optimal solutions associated with  $\widehat{\bS}^\star$ and all the $\widehat{\bS}_i$. Now, let $\left(\widehat{\bS}_{i'},\widehat{\bM}_{i'} \right)$ be a sequence extracted from $\left(\widehat{\bS}_{i},\widehat{\bM}_{i} \right)$ that converges to a point $ \left(\widehat{\bS}^\star,\widehat{\bM}_1^\star\right)$. Due to the continuity of $\|\cdot\|$, $\|\widehat{\bS}^\star - \widehat{\bM}_1^\star\|=\displaystyle{\lim_{i' \rightarrow \infty}}\displaystyle{\min_{\widehat{\bM}\in \mathcal{M}}}\|\widehat{\bS}_{i'} - \widehat{\bM}\|=\displaystyle{\min_{\widehat{\bM}\in \mathcal{M}}}\|\widehat{\bS}^\star - \widehat{\bM}\|$, namely $\widehat{\bM}_1^\star=\widehat{\bM}^\star$ is the optimal solution. Finally, if $\widehat{\bM}_{i}$ does not converge to $\widehat{\bM}^\star$ there exists at least one extract sequence $\widehat{\bM}_{i_1'}$ such that $\widehat{\bM}_{i_1'}\rightarrow\widehat{\bM}_2^\star\neq \widehat{\bM}^\star$,  which is an absurd since the optimal solution is unique. 
\end{proof}

\section*{APPENDIX S2. Proof of the equivalence between problems ${\cal{P}}$ and ${\cal{P}}_1$}

\begin{proof}
Let $\bar{\bM}$ be a feasible solution to ${\cal{P}}$ and $\left(\bar{\bR},\bar{\sigma}_n^2\right)$ be such that $\left(\bar{\bM},\bar{\bR},\bar{\sigma}_n^2\right)$ satisfies the constraints in (3). Since $\left(\bar{\bM},\bar{\bR}+(\bar{\sigma}_n^2-{\sigma^2})\bI,{\sigma^2}\right)$ is also feasible to (3), ${\cal{P}}$ is equivalent to
\begin{equation}
{\cal{P}^{\prime}}\left\{
\begin{array}[c]{ll}
\underset{\bM}{\min} &
\begin{array}[c]{c}
\|\bM - \widehat{\bS}\|
\end{array}
\\
\mbox{s.t.} &
\begin{array}[t]{l}
\frac{\lambda_{max}\left(\bM\right)}{\lambda_{min}\left(\bM\right)}\leq \kappa_M \\
\sigma^2\bI+\bR=\bM\\
\bR\succeq {\bf 0}
\end{array}
\end{array} \right..
\end{equation}
Next, observe that the set
$$
\left\{
\begin{array}[c]{ll}
\sigma^2\bI+\bR=\bM\\
\bR\succeq {\bf 0}
\end{array}
\right.,
$$
can be recast as
$$
\left\{
\begin{array}[c]{l}
\bR=\bM-\sigma^2\bI\\
\bM\succeq\sigma^2 \bI
\end{array} \right..
$$
Hence, defining $\bX= \frac{\bM}{\sigma^2}$, ${\cal{P}^{\prime}}$ boils down to 
\begin{equation}
{\cal{P}}_1\left\{
\begin{array}[c]{ll}
\underset{\bX}{\min} &
\begin{array}[c]{c}
\|\bX - \bS\|
\end{array}
\\
\mbox{s.t.} &
\begin{array}[t]{l}
\bX\succeq \bI\\
\frac{\lambda_{max}\left(\bX\right)}{\lambda_{min}\left(\bX\right)}\leq \kappa_M \\
\end{array}
\end{array} \right.,
\end{equation}
where $\bS =\frac{\widehat{\bS}}{\sigma^2}$. Since $\frac{\lambda_{max}(\bX)}{\lambda_{min}(\bX)}$ is quasi-convex function over the set $\bX\succeq\bI$, ${\cal{P}}_1$ is a convex problem.
Finally, notice that Problem ${\cal{P}}_1$ is equivalent to 
\begin{equation}
{\cal{P}}_1^{\prime}\left\{
\begin{array}[c]{ll}
\underset{\bX}{\min} & \|\bX - \bS\|\\
\mbox{s.t.} & \bX\succeq \bI\\
 & \|\bX - \bS\|\preceq\|\bI - \bS\|\\
 & \frac{\lambda_{max}(\bX)}{\lambda_{min}(\bX)}\leq{\kappa_M }
\end{array}
\right..
\end{equation}
Now, since the objective in ${\cal{P}}_1^{\prime}$ is a continuous function  and the feasible set is a compact set,  Weierstrass theorem ensures the existence of a feasible point $\bX^\star$ to ${\cal{P}}_1^{\prime}$ such that $v({\cal{P}}_1)=v({\cal{P}}_1^{\prime})=\|\bX^\star - \bS\|$, which concludes the proof.
\end{proof}

\section*{APPENDIX S3. Proof of the equivalence between ${\cal{P}}_2$ and ${\cal{P}}_2^{\prime}(u)$}
\begin{proof}
Let us observe that the constraint set
\begin{equation}\label{vincolo_senza_u}
\left\{
\begin{array}[c]{l}
\bLambda\succeq\bI\\
\frac{\lambda_{max}(\bLambda)}{\lambda_{min}(\bLambda)}\leq{\kappa_M }\\
{\bLambda}=\diag\left([\lambda_1,\ldots,\lambda_N]^T\right)
\end{array}\right.,
\end{equation}
is equivalent to
\begin{equation}\label{vincolo_senza_u_new}
\left\{
\begin{array}[c]{l}
\lambda_i\geq 1, i=1,\ldots, N\\
\lambda_i\leq {\kappa_M } \displaystyle{\min_h}\{\lambda_h\}, i=1,\ldots, N
\end{array}\right..
\end{equation}
Now, introducing an auxiliary variable $u>0$, the set (\ref{vincolo_senza_u_new}) can be cast as
\begin{equation}\label{vincolo_senza_u_new1}
\left\{
\begin{array}[c]{l}
\lambda_i\geq 1, i=1,\ldots, N\\
u\leq{\lambda_i}\leq \kappa_M u, i=1,\ldots, N\\
u>0
\end{array}\right..
\end{equation}
Indeed, if $\bar{\blambda}$ is a feasible point to (\ref{vincolo_senza_u_new}), 
$\left(\bar{\blambda},\min\left\{\bar{\blambda}\right\}\right)$ is a feasible point to (\ref{vincolo_senza_u_new1}). On the other hand, if $\left(\bar{\blambda}^1,\bar{u}^1\right)$ is a feasible point to (\ref{vincolo_senza_u_new1}), 
$$0< \frac{\bar{\lambda}^1_i}{\displaystyle{\min_h}\left\{\bar{\lambda}_h^1\right\}}\leq \frac{\bar{u}^1\kappa_M}{\displaystyle{\min_h}\left\{\bar{\lambda}_h^1\right\}}\leq  \frac{\bar{u}^1 \kappa_M}{\bar{u}^1}=\kappa_M.$$
As a result, $\bar{\blambda}_1$ is a feasible point to (\ref{vincolo_senza_u_new}).
Finally, since (\ref{vincolo_senza_u_new1}) is empty if $u<\frac{1}{\kappa_M }$, ${\cal{P}}_2^{\prime}(u)$  follows.
\end{proof}

\section*{APPENDIX S4. Proof of the results in equations (16) and (17)}\begin{proof}
To prove equations (16) and (17), let us introduce the following notation
$$
\alpha_1 = \max(1,u),\quad \beta_1 = \max(d_i,\alpha).$$
Hence, $\lambda_i(u) = \min(\kappa_M u,\beta_1)$.
Now, assuming $d_i > 1$:
\begin{itemize}
\item if $\frac{1}{\kappa_M } \leq u < \frac{d_i}{\kappa_M }$, then $u < d_i$ and $\beta_1 = d_i$. Moreover,
$\quad u \kappa_M  < d_i$, implying that $\lambda_i(u) = u \kappa_M $ as well as $h_i(u) = |u \kappa_M  - d_i| = d_i - u \kappa_M $;
\item if $\frac{d_i}{\kappa_M } \leq u < d_i$, as in the previous item $\beta_1 = d_i$. Now,
${u\kappa_M }\geq{d_i}$, and thus $\lambda_i(u) = d_i$, i.e., $h_i(u) = 0$;
\item if $u\geq d_i$, then $\alpha_1=u$ and $\beta_1=u$; thus, $\lambda_i(u)=u$ and $h_i(u)=u-d_i$.
\end{itemize}
\noindent
Instead, when $d_i\leq 1$:
\begin{itemize}
\item if $\frac{1}{\kappa_M }\leq u < 1$, then $\alpha_1=1$ and $\beta_1=1$; hence, $\lambda_i(u)=1$ and $h_i(u) = 1 - d_i$;
\item if $u\geq 1$, then $\beta_1=u$; thus, $\lambda_i(u) = u$ and $h_i(u)=u-d_i$.
\end{itemize}
\end{proof}

\section*{APPENDIX S5. Proof of the claims concerning $G_1(u)$}
\begin{proof}
Let us start assuming  $d_1\leq 1$. This implies that $d_i\leq 1$, $i=1,\ldots,N$  and
\begin{equation}\label{frob_funzione_h_autoval_minore_di_1}
(h_i(u))^2 = \left\{
\begin{array}[c]{ll}
(1 - d_i)^2 & \mbox{if}\quad \frac{1}{\kappa_M }\leq u < 1\\
(u - d_i)^2 & \mbox{if}\quad u \geq 1\\
\end{array}
\right..
\end{equation}
Consequently, for $i=1,\ldots,N$, each $G_1(u)$ is a constant function over $u\in\left[\frac{1}{\kappa_M },1\right[$ and it is monotonically increasing in $u\in\left[1,+\infty\right]$. Thus, $u^\star=\frac{1}{\kappa_M}$.

Consider now $1 < d_1 \leq \kappa_M $. Let $\mathcal{I}_p=\{i:d_i>1\}$ be the set of indexes corresponding to the eigenvalues greater than $1$. Since $\frac{d_1}{\kappa_M }\leq 1< d_{p_{max}}$, where $p_{max}\in\mathcal{I}_p$ is the index associated with the lowest eigenvalue greater than 1, 
\begin{equation}\label{eq_hi}
\displaystyle{\sum_{i\in I}} (h_i(u))^2
\end{equation}
is monotonically decreasing over $\frac{1}{\kappa_M }\leq u \leq \frac{d_1}{\kappa_M}$. Moreover, (\ref{eq_hi}) is constant over $\frac{d_1}{\kappa_M }\leq u \leq d_{p_{max}}$ and monotonically increases for $u\geq d_{p_{max}}$. Furthermore, 
if $d_i\leq 1$, $(h_i(u))^2$ is a constant function over $\frac{1}{\kappa_M }\leq u<1$ and monotonically increases over $u\geq 1$. Hence, $u^\star=\frac{d_1}{\kappa_M }$.

Finally, if $d_1 > \kappa_M $ $G_1(u)$ is a strictly decreasing function over $\frac{1}{\kappa_M } \leq u < 1$, whereas it is a strictly increasing function within $d_1 \leq u < +\infty$. Hence, any minimum belongs to the interval $\left[1, d_1\right]$.

Let us now focus on the differentiability of $G_1(u)$. If $d_i\leq 1$, the derivative of  $(h_i(u))^2$ is
\begin{equation}
\frac{d (h_i(u))^2}{du} = \left\{
\begin{array}[c]{ll}
0 & \mbox{if}\quad \frac{1}{\kappa_M }\leq u < 1\\
2(u - d_i) & \mbox{if}\quad u > 1\\
\end{array}
\right..
\end{equation}
Hence,  $\frac{d (h_i(u))^2}{du}$ is continuous over  $u \geq 1$ where in $u=1$  the right derivative is considered. If $d_i > 1$, \begin{equation}
\frac{d (h_i(u))^2}{du} = \left\{
\begin{array}[c]{ll}
2(u \kappa_M ^2 - d_i \kappa_M ) & \mbox{if}\quad \frac{1}{\kappa_M }\leq u < \frac{d_i}{\kappa_M }\\
0 & \mbox{if}\quad \frac{d_i}{\kappa_M } < u < d_i\\
2(u - d_i) & \mbox{if}\quad u > d_i
\end{array}
\right..
\end{equation}
Hence, in each of the sub-intervals $\left[\frac{1}{\kappa_M },\frac{d_i}{\kappa_M }\right[$, $\left]\frac{d_i}{\kappa_M },d_i\right[$, and $\left]d_i,+\infty\right[$, $\frac{d (h_i(u))^2}{du}$ is continuous; moreover, in correspondence of the points $u = \frac{d_i}{\kappa_M }$ and $u = d_i$, the right and left derivatives coincide implying that the overall derivative is continuous over $u\geq \frac{1}{\kappa_M }$. Since $\frac{1}{\kappa_M }\leq 1$, the function $\frac{d G_1(u)}{du} = \displaystyle{\sum_{i=1}^{N}} \frac{d(h_i(u))^2}{du}$, is continuous over $u\in\left[1,d_1\right]$.

As to the convexity of $G_1(u)$, it easily follows from the convexity of each term.
\end{proof}

\end{document}